\documentclass[aps,prd,amsmath,floats,floatfix,
  nofootinbib,
  notitlepage,
  showpacs]{revtex4-1}

\usepackage{amssymb}
\usepackage{amsmath}
\usepackage{verbatim}
\usepackage{mathrsfs}
\usepackage{amsfonts}
\usepackage{latexsym}
\usepackage{epsfig}
\usepackage{color}
\usepackage{graphicx}
\usepackage{units}
\usepackage{overpic}
\usepackage[hidelinks]{hyperref}
\usepackage{afterpage}
\usepackage{float}
\usepackage{mathtools}
\usepackage{xcolor}
\hypersetup{
    colorlinks,
    linkcolor={red!50!black},
    citecolor={blue!50!black},
    urlcolor={blue!80!black}
}

\usepackage{booktabs, multirow} 
\usepackage{soul}
\usepackage{changepage,threeparttable} 
\begin{document}


\definecolor{orange}{rgb}{0.9,0.45,0} 
\definecolor{applegreen}{rgb}{0.55, 0.71, 0.0}


\title{Magnetostatic boson stars}

\author{V\'ictor Jaramillo}
\affiliation{Instituto de Ciencias Nucleares, Universidad Nacional Aut\'onoma de M\'exico,
Circuito Exterior C.U., A.P. 70-543, M\'exico D.F. 04510, M\'exico}

\author{Dar\'io N\'u\~nez}
\affiliation{Instituto de Ciencias Nucleares, Universidad Nacional Aut\'onoma de M\'exico,
Circuito Exterior C.U., A.P. 70-543, M\'exico D.F. 04510, M\'exico}


\date{\today}


\begin{abstract}
  We solve the Einstein-Maxwell-Klein-Gordon system of equations
  and derive a compact, static axially symmetric magnetized object
  which is electrically neutral
  and made of two complex massive charged scalar fields.
  We describe several properties of such solution, including the
  torus form of the matter density and the expected dipolar distribution
  of the magnetic field, with some peculiar features in the central regions.
  The solution shows no divergencies in any of the field and metric functions.
  A discussion is presented on 
  a case where the
  gravitational and magnetic fields in the external region are similar to those of neutron stars. 
\end{abstract}


\pacs{
04.20.-q, 
04.25.Dm, 
95.30.Sf 
}


\maketitle


\section{Introduction}
\label{Sec:intro}

Boson stars are self-gravitating solitons made up of complex scalar field.
This objects are interesting for various reasons.
They serve as simple models for compact objects in the interplay
of Field Theory and General Relativity
and are also interesting on their own
since they possess important dynamical properties
that allow to study them in hypothetical strong gravity astrophysical scenarios,
for example, in the gravitational waveform research;
see \textit{e.g.}~\cite{Bustillo:2020syj}.
Bosonic stars have applications also as
black hole mimickers \cite{Guzman:2009zz,Herdeiro:2021lwl} and, more generally,
scalar fields are relevant in cosmology as quintessence \cite{Tsujikawa:2013fta},
ultralight dark matter \cite{Suarez:2013iw, Hui:2016ltb, Urena-Lopez:2019kud}
and as a source of inflation \cite{Guth:1980zm}. 

Static and stationary single field configurations have been presented in
the literature in the past years (for reviews see
Refs.~\cite{Liebling:2012fv,Visinelli:2021uve,Shnir:2022lba}),
for instance, the free and massive bosonic field solution
of the Einstein-Klein-Gordon equations
in spherical symmetry \cite{Kaup:1968zz} can be generalized gauging the
global $U(1)$ symmetry, which leads to the charged version of boson stars
\cite{Jetzer:1989av}. Rotating generalizations of boson stars were obtained
in \cite{Yoshida:1997qf} within the Einstein-Klein-Gordon setup and
in the Einstein-Klein-Gordon-Maxwell extension, the charged rotating one
\cite{Collodel:2019ohy}.
In these models the coupling constant parameter is freely specifiable,
however, in order to obtain
equilibrium solutions, the value of the coupling parameter ranges from zero
up to a critical value \cite{Jetzer:1989av,Collodel:2019ohy}.

Staying within the complex, massive, free scalar field case there are
two more characterizations that can be found in the literature
up to the present time,
these are
the multipolar boson stars \cite{Herdeiro:2020kvf}, which are static nonspherical
configurations with similar morphologies to the probability density
of atomic orbitals and the multifield ($\ell$-) boson stars
\cite{Alcubierre:2018ahf,Jaramillo:2020rsv,Sanchis-Gual:2021edp}, in which 
the $U(1)$ symmetry is generalized by considering a $U(N)$ symmetry.
Of particular interest in this paper is
the toroidal static boson star of \cite{Sanchis-Gual:2021edp},
which can be understood as the superposition of two contrarotating solutions
that give rise to a static equilibrium configuration.

The rotating charged boson stars share some properties with the uncharged
rotating case, such as the toroidal shape and with the charged (electrostatic)
case, such as the critical value of the coupling constant. As one would expect
the charged and rotating  general solutions include electric
charge and magnetic dipole moment with the particularity
that the magnetic moment is nonzero only if the electric charge is nonzero,
therefore obtaining solutions where both electric and magnetic fields are
present in the local inertial frame of zero angular momentum.
Until now, no electrically neutral and magnetized
self-gravitating bosonic stars have been
constructed, which might be relevant models
in the study of strong gravity and magnetic fields phenomena.

Magnetic fields play an important role in many astrophysical scenarios.
Some of the relativistic applications involve compact, electrically neutral objects and
strong magnetic fields where their
self-gravitation must be taken into account. Fully relativistic and
self-consistent models of neutron stars with magnetic fields
where first presented in \cite{Bocquet:1995je} (see also \textit{e.g.}
\cite{Cardall:2000bs,Chatterjee:2014qsa} for poloidal
and \cite{Oron:2002gs,Kiuchi:2008ch,Frieben:2012dz} for toroidal
magnetic fields). These are numerical solutions of the
Einstein-Maxwell-Euler system in axial symmetry which can possess
angular momentum and total electric charge.
Even the globally neutral and static cases are deformed by the effect of the
magnetic field, changing in consequence the global properties of
the equilibrium configurations.

The purpose of this paper is to construct and study
magnetostatic solutions of boson stars with zero total electric
charge. To do so we obtain a generalization of the toroidal
static boson stars by coupling the scalar fields to the
electromagnetic field, we shall show that the gauge coupling
parameter can exceed the critical value obtained for the
electrostatic boson stars. In Sec.~\ref{Sec:model} we present
the action for the model, the ans\"atze for the fields and
the conserved quantities. Then, in Sec.~\ref{Sec:solutions} we
describe the numerical procedure and the construct sequence of
solutions with different values of the coupling parameter, discussing
the physical properties of the configurations.
The magnetic field for the constructed solutions
and a comparison with strongly magnetized neutron stars
is presented Sec.~\ref{Sec:magnetic}.
We conclude our manuscript and give some perspectives of future works
in Sec.~\ref{Sec:Conclusions}. In Appendix~\ref{Sec:3+1}
we give the complete set of elliptic partial differential equations
of our model together with explicit expressions for the $3+1$
decomposition of the energy momentum tensor. We work with $c=G=1$
and the metric signature is taken to be $(-,+,+,+)$.

\section{Model}
\label{Sec:model}

A single electrically charged scalar field in spherical symmetry allows
to construct charged boson stars, which are
static solutions that give rise to an electric field
as measured by an observer at rest \cite{Jetzer:1989av}. Even the rotating
generalization of this configurations,
which also generate a magnetic field, possess an electric field that does not
vanish \cite{Collodel:2019ohy}. There are no immediate simple models
consisting of one scalar field that give rise to
magnetostatic self-gravitating solutions, however the multifield approach
leads to a natural way of obtaining such boson stars.

On the other hand, although fully relativistic neutron stars with
magnetic fields have been constructed,
all the solutions (to the best of our knowledge)
have been obtained using the free current assumption,
i.e., electromagnetic
sources $J^\mu$ independent to the fluid movement.
This means in particular that in the (magneto)static
solutions the fluid is at rest while
the spatial components of the electric current are nonzero. This is a limiting
assumption, and as pointed out
in \cite{Chatterjee:2014qsa}, in principle the currents should be derived from a microscopic
model which``would require a multifluid approach to model the movements of free
protons and electrons''.

For boson stars, in the Einstein-Klein-Gordon-Maxwell framework, the free current assumption
cannot be
even made since the electric current is determined by the scalar fields, however the
multifield approach is feasible. In our approach globally neutral configurations
are constructed by superposition of two contrarotating ``thick current loops''
made of charged scalar fields.

The general framework in which magnetized boson stars are constructed
consists of two self-gravitating complex scalar fields minimally coupled
to the electromagnetic four-potential with coupling constants of opposite
sign. In this section we summarize the basic equations needed to construct
the solutions and the conserved quantities that will be useful in the analysis.

\subsection{Field equations}
We consider two massive complex scalar fields, $\Phi_{(1)}$ and $\Phi_{(2)}$,
minimally coupled to the Einsteinian gravity and to the electromagnetic field,

\begin{equation}\label{eq:action}
S=\int d^4 x\sqrt{-g} \left[\frac{1}{16\pi }R-\frac{1}{2}\sum_{j=1}^2\left(g^{\mu\nu}(D^{(j)}_\mu\Phi_{(j)})(D^{(j)}_\nu\Phi_{(j)})^*+\mu^2|\Phi_{(j)}|^2\right)-\frac{1}{4\mu_0}F_{\mu\nu}F^{\mu\nu}\right],
\end{equation}
where $F_{\mu\nu}=\partial_\mu A_\nu-\partial_\nu A_\mu$ is the Faraday tensor
and the covariant derivative operators, $D^{(1)}_\mu=\nabla_\mu+iqA_\mu$ and
$D^{(2)}_\mu=\nabla_\mu-iqA_\mu$ couple both scalar fields with $A_\mu$.
Notice that we have chosen equal mass terms for both scalar fields and opposite
signs for the electromagnetic coupling constants (boson charges). The scalar fields
interact with each other indirectly, through gravity and the electromagnetic field.

Variation of Eq.~(\ref{eq:action}),
with respect to the different fields leads to the Euler-Lagrange equations
of the model (see \textit{e.g.} \cite{Hawking:1973uf}).
Variation with respect to $g_{\mu\nu}$ leads to
\begin{subequations}\label{eq:ekgm}
\begin{eqnarray}
&& R_{\mu\nu}-\frac{1}{2}R g_{\mu\nu}=8\pi T_{\mu\nu} ;
\label{eq:einstein} \\
&& T_{\mu\nu}=T_{\mu\nu}^{(1)}+T_{\mu\nu}^{(2)}+(T^{\mathrm{EM}})_{\mu\nu} ,
\label{Eq:tmunu}\\
&& T_{\mu\nu}^{(j)}:=\frac{1}{2}(D^{(j)}_{\mu}\Phi_{(j)})(D^{(j)}_{\nu}\Phi_{(j)})^*+\frac{1}{2}(D^{(j)}_{\nu}\Phi_{(j)})(D^{(j)}_{\mu}\Phi_{(j)})^*
-\frac{1}{2}g_{\mu\nu}\left(g^{\alpha\beta}(D^{(j)}_{\alpha}\Phi_{(j)})(D^{(j)}_{\beta}\Phi_{(j)})^*+\mu^2|\Phi_{(j)}|^2\right),\\
&& (T^{\mathrm{EM}})_{\mu\nu}:=\frac{1}{\mu_0}F_{\mu\sigma} F_{\nu\lambda}g^{\sigma\lambda}-\frac{1}{4\mu_0} g_{\mu\nu} F_{\alpha\beta} F^{\alpha\beta}.
\end{eqnarray}
\end{subequations}

The equation for the fields $\Phi_{(1)}$ and $\Phi_{(2)}$ are
the Klein-Gordon equations,
\begin{equation}\label{eq:kg}
    g^{\mu\nu}D_\nu^{(j)} D_\mu^{(j)} \Phi_{(j)}=\mu^2\Phi_{(j)} .
\end{equation}

Variation with respect to $A_\mu$ leads to the Maxwell equations with
source the charged scalar fields which define a current four-vector $J^\mu$,
\begin{subequations}\label{eq:maxwell}
  \begin{eqnarray}
    &&\nabla_\nu F^{\mu\nu}=\mu_0J^\mu:=\mu_0(q j^\mu_1-q j^\mu_2);\\
&&j^\mu_i:=\frac{ig^{\mu\nu}}{2}({\Phi_{(i)}}^*D_\nu^{(i)}\Phi_{(i)}-\Phi_{(i)}(D_\nu^{(i)}{\Phi_{(i)}})^*),
  \end{eqnarray}
\end{subequations}
here $J^\mu$ is the (total) electromagnetic current.

\subsection{Global quantities}
The spacetime we will consider in this work is stationary (static) and axisymmetric.
Komar expressions allow to calculate global quantities for each of this isometries;
if $\xi$ is the Killing vector associated with stationarity,
$\Sigma_t$ is a spacelike surface and $n^\mu$ the unit vector normal to this hypersurface,
then the quantity,
\begin{equation}\label{eq:KomarM}
  M=\frac{1}{4\pi}\int_{\Sigma_t}R_{\mu\nu}n^\mu\xi^\nu dV,
\end{equation}
defines the Komar mass. Similarly, if $\chi$ is the Killing vector
associated with the axial symmetry, the quantity

\begin{equation}
  J=\frac{1}{8\pi}\int_{\Sigma_t}R_{\mu\nu}n^\mu\chi^\nu.
\end{equation}
gives the angular momentum of the spacetime. 

The quantities $j^\mu_1$ and $j^\mu_2$ defined in Eq.~(\ref{eq:maxwell})
are Noether density currents ($\nabla_\mu j^\mu=0$) which arise
from the invariance of
Eq.~(\ref{eq:action}) under the $U(1)$ gauge transformation of
$\Phi_{(1)}$ and $\Phi_{(2)}$.
It follows that integration of the projection onto $n^\mu$ of this currents
over $\Sigma_t$ leads to the conserved particle numbers
\begin{equation}
  \mathcal{N}_1=\int_{\Sigma_t} j^\mu_1 n_\mu dV, \quad \mathcal{N}_2=\int_{\Sigma_t} j^\mu_2 n_\mu dV; \quad \mathcal{N}:=\mathcal{N}_1+\mathcal{N}_2.
\end{equation}

In the rotating boson stars, it was shown
\cite{Yoshida:1997qf,Grandclement:2014msa,Ontanon:2021hbg} that the
angular momentum $J$ takes values that are integer multiples of the particle number, 
$J=m\mathcal{N}$, with $m$ the
winding number (defined below) of the scalar field ansatz, this result is
 also valid in the charged rotating case \cite{Collodel:2019ohy}.
 However, in the magnetized solutions obtained in this work this relation 
 does not hold since they are by construction
 $J=0$ static, as will be argued in the next section.

The associated total electric charge, related to the sources 
at the Maxwell equation is given by
$Q=\int_{\Sigma_t} J^\mu n_\mu dV =q(\mathcal{N}_1- \mathcal{N}_2).$
In the single field static and rotating charged boson stars the
total charge of the system is related to the particle number by $Q=q\mathcal{N}$ 
and it was obtained \cite{Jetzer:1989av} that $Q$ coincides with the asymptotic value
extracted from the electric potential and matches the exterior
Reissner-Nordstr\"om solution. Again, the relation $Q=q\mathcal{N}$ is not
valid for our case because, as we will see bellow, the obtained 
solutions satisfy $Q=0$.

\subsection{Static axisymmetric spacetime and Ans\"atze for the fields}

In coordinates adapted to the Killing fields, where $\xi=\partial/\partial t$
and $\chi=\partial/\partial \varphi$, the general static and axially symmetric line
element we will consider is in the Lewis-Papetrou form,
\begin{equation}\label{eq:metric}
  g_{\mu\nu}dx^\mu dx^\nu=-e^{2 F_0} dt^2 + e^{2 F_1}(dr^2+r^2 d\theta^2) + e^{2F_2}r^2\sin^2\theta d\varphi^2,
\end{equation}
where the metric functions $F_0$, $F_1$ and $F_2$ depend only on
the coordinates $r$ and $\theta$. We have used the same line element 
as the one in Ref.~\cite{Sanchis-Gual:2021edp}, where the
toroidal static boson star is constructed,
however, the $g_{t\varphi}$ term usually written as the function
$w(r,\theta)$ or $w(r,\theta)/r$, is not included in (\ref{eq:metric})
because we are looking
for static configurations with zero total angular momentum $J$, and in this case
it can be seen \cite{Stephani:2003tm,Gourgoulhon:2010ju} that $w=0$ if
and only if the spacetime is static. 

The contribution of the scalar fields in the energy-momentum tensor,
$T_{\mu\nu}^{(1)}$, $T_{\mu\nu}^{(2)}$ will be consistently independent
of $t$ and $\varphi$ if for the scalar fields we use the 
following ansatz, which is similar to the one
 used for rotating, multifield, multifrequency boson stars and even for chains \cite{Herdeiro:2021mol},

\begin{equation}\label{eq:ansatzphi}
\Phi_{(1)}=\phi(r,\theta)e^{i\omega t - i m\varphi};\qquad \Phi_{(2)}=\phi(r,\theta)e^{i\omega t + i m\varphi}.
\end{equation}

Here $m$ is and integer called winding number. Moreover, the opposite sign of
this parameter for each field in Eq.~(\ref{eq:ansatzphi}) can be interpreted
as having contra-rotating scalar fields distributions. It is not difficult to
obtain that with this election of winding numbers, $T_{t\varphi}^{(1)}=-T_{t\varphi}^{(2)}$,
consistent with the Einstein tensor component $G_{t\varphi}$ being zero for
the metric in Eq.~(\ref{eq:metric}).

Again, analyzing the components of the Einstein tensor
we can elucidate the anzatz for the field $A_\mu$. Two possibilities for the
electromagnetic four-potential are compatible with the spacetime at hand:
the purely poloidal ($A_r=A_\theta=0$) and the purely toroidal
($A_t=A_\varphi=0$) magnetic fields\footnote{In both cases, the circularity
property of spacetime is not broken and the metric tensor takes the form (\ref{eq:metric}).}
\cite{Kiuchi:2008ch, Oron:2002gs}, however only the first possibility
can be realized given the ansatz~(\ref{eq:ansatzphi}) chosen for the
scalar fields since only the $J^\varphi$ source of the Maxwell equations is
nonzero\footnote{In the single field charged rotating star, also the $J^t$
component is nonzero, in our case however $j_1^t=j_2^t$, see Appendix~\ref{Sec:3+1}.},
which additionally implies that $A_t=\mathrm{constant}$, therefore we adopt
\begin{equation}\label{eq:ansatzA}
  A_\mu dx^\mu=C(r,\theta) d\varphi.
\end{equation}

The resulting number density currents, given in Eqs.~(\ref{eq:j1}) and (\ref{eq:j1}), imply $Q=0$ since $J^\mu n_\mu=0$.

\section{Numerical solutions}
\label{Sec:solutions}

\subsection{Boundary conditions and numerical method}
In order to construct magnetostatic solutions of boson stars,
the Einstein-Klein-Gordon-Maxwell system is solved. This means
solving for the five functions $\{\phi,C,F_0,F_1,F_2\}$
and the unknown parameter $\omega$,
imposing the appropriate symmetries and boundary conditions. The
full elliptic system of coupled partial differential equations
(PDEs)
in $r$ and $\theta$ is given in the Appendix~\ref{Sec:3+1}.

First, we impose even parity
with respect to reflections at the equatorial
plane of the five unknown functions which in particular implies that
derivatives with respect to $\theta$ at $\theta=\pi/2$ vanish and
also that the
required integration domain reduces to $0\leq\theta\leq\pi/2$,
$0\leq r<\infty$.

Asymptotic flatness implies that the following outer boundary
conditions must be imposed,
\begin{equation}\label{eq:out_bc}
  \begin{split}
    &\phi|_{r\to\infty}=0,\quad C|_{r\to\infty}=0;\\
    &F_0|_{r\to\infty}=0, \quad F_1|_{r\to\infty}=0, \quad F_2|_{r\to\infty}=0.
  \end{split}
\end{equation}

Also the condition $\omega<\mu$ is necessary in order to have
$\phi|_{r\to\infty}=0$.
Regularity of the solution at the origin and on the symmetry axis require,

\begin{equation}\label{eq:regularity_r}
  \begin{split}
    &\phi|_{r=0}=0, \quad C|_{r= 0}=0;\\
    &\partial_r F_0 |_{r=0}=0, \quad \partial_r F_1 |_{r=0}=0, \quad \partial_r F_2 |_{r=0}=0,\\
    &F_1|_{r=0}=F_2 |_{r=0}.
  \end{split}
\end{equation}
\begin{equation}\label{eq:regularity_th}
  \begin{split}
    &\phi|_{\theta=0,\pi}=0, \quad C|_{\theta= 0,\pi}=0;\\
    &\partial_\theta F_0 |_{\theta=0,\pi}=0, \quad \partial_\theta F_1 |_{\theta=0,\pi}=0, \quad \partial_\theta F_2 |_{\theta=0,\pi}=0,\\
    &F_1|_{\theta=0,\pi}=F_2 |_{\theta=0,\pi}.
  \end{split}
\end{equation}
The regularity conditions for $\phi$ for the case $m=0$ are different
from those of the previous expressions, however this case
reduce to
the widely studied spherical, nonrotating, neutral boson star,
and will not be addressed in this manuscript except for comparison.

The nonlinear PDEs are solved numerically using the spectral solver
\textsc{Kadath} \cite{Grandclement:2009ju,Kadath}
which implements a Newton-Raphson iteration.
This library, which has been
successfully applied to solve a wide variety of PDEs in theoretical
physics and in particular in relativity,
was also used in the construction of rotating
boson stars \cite{Grandclement:2014msa}.

Chebyshev polynomials have been used in the spectral method
as basis functions for the expansions
of the five unknown functions.
The spatial domain is divided into 8 spherical shells with boundaries located
at $r=\{2,4,8,16,32,64,128\}$.
The regularity conditions in Eqs.~(\ref{eq:regularity_r}) and (\ref{eq:regularity_th})
are either imposed by the spectral
basis\footnote{Details on how this is implemented in terms of the Chebyshev spectral basis
in the innermost shell and on the
symmetry axis for similar problems can be found in \cite{Grandclement:2014msa} and \cite{Grandclement:2007sb}.}
for a given function on the corresponding domain,
or checked that they hold up to numerical accuracy. On the other hand
the outer boundary conditions in Eq.~(\ref{eq:out_bc}) are imposed
``exactly'' (without the need of a cutoff radius) given the compactification
of the radial variable at the outermost spherical shell.

An initial guess for the functions is required in order to start
the iteration. For each value of $m$ this needs to be done only once.
The expressions
\begin{equation}
  N:=e^{F_0}=1-(1-N_0)e^{-r^2/r_0^2},\quad F_1=F_2=0,\quad C=0,
\end{equation}
also used
in \cite{Alcubierre:2021psa}, and
\begin{equation}
  \phi=\phi_0 (r\sin\theta)^m e^{-(x^2/2+2 z^2)m/r_0^2}
\end{equation}
with $x=r\sin\theta$, $z=r\cos\theta$, proposed in Ref.~\cite{Grandclement:2014msa}, have proven
to be good
guesses given certain choice of $\phi_0$, $r_0$ fixing
the coupling constant $q=0$ and the lapse at $r=0$, $N_0\approx0.95$.
The last condition prevents convergence to the trivial $\phi=0$ solution
and also leads to a Newtonian configuration ($\omega\sim\mu$).
Once the first solution is obtained the rest of the solutions are obtained
by varying $N_0$ and increasing $q$ by small steps.

In addition to the Komar expression on Eq.~(\ref{eq:KomarM}), the ADM mass
definition can be used to obtain the total mass of the star.
Both quantities should coincide given the stationarity
of the spacetime
we are considering \cite{Gourgoulhon:2010ju}, therefore the difference between 
the ADM and the Komar masses can be used as an indicator of
the numerical accuracy and provide an estimation of the
numerical error of the solution.
An expression for the ADM mass suitable for our case is the
expression~\cite{Gourgoulhon:2010ju},
\begin{equation}
  M=-\frac{1}{8\pi}\lim_{S\to\infty}\oint_S\left[\frac{\partial}{\partial r}(F_1+F_2)+\frac{F_2-F_1}{r}\right]r^2\sin\theta d\theta d\varphi,
\end{equation}
where the limit indicates integration over a sphere $S$ of radius $r\to\infty$.

One can also verify that the value of the unknown frequency $\omega$
converge exponentially to a finite value with the number of collocation points.
This error indicator has been used together with the relative difference of
the ADM and Komar masses to monitor accuracy along the sequence of numerical solutions
and to carry out convergence tests of the
solutions with increasing number of spectral coefficients.

\subsection{Structure of the stars}
\label{Sec:structure}

One can use the code to solve for boson stars of several types.
The spectral code has been able to reproduce sequences of
solutions already presented in the literature such as the
single field static and the rotating mini-boson stars, as well as the multifield
$\ell$-boson stars and the toroidal static boson stars. In this section
we present new solutions that correspond to magnetostatic boson stars. 
These configurations generalize the toroidal static boson stars,
incorporating a new parameter, $q$,
in addition to the frequency $\omega$ and the winding number $m$.

\begin{figure}
  \includegraphics[width=0.3\textwidth]{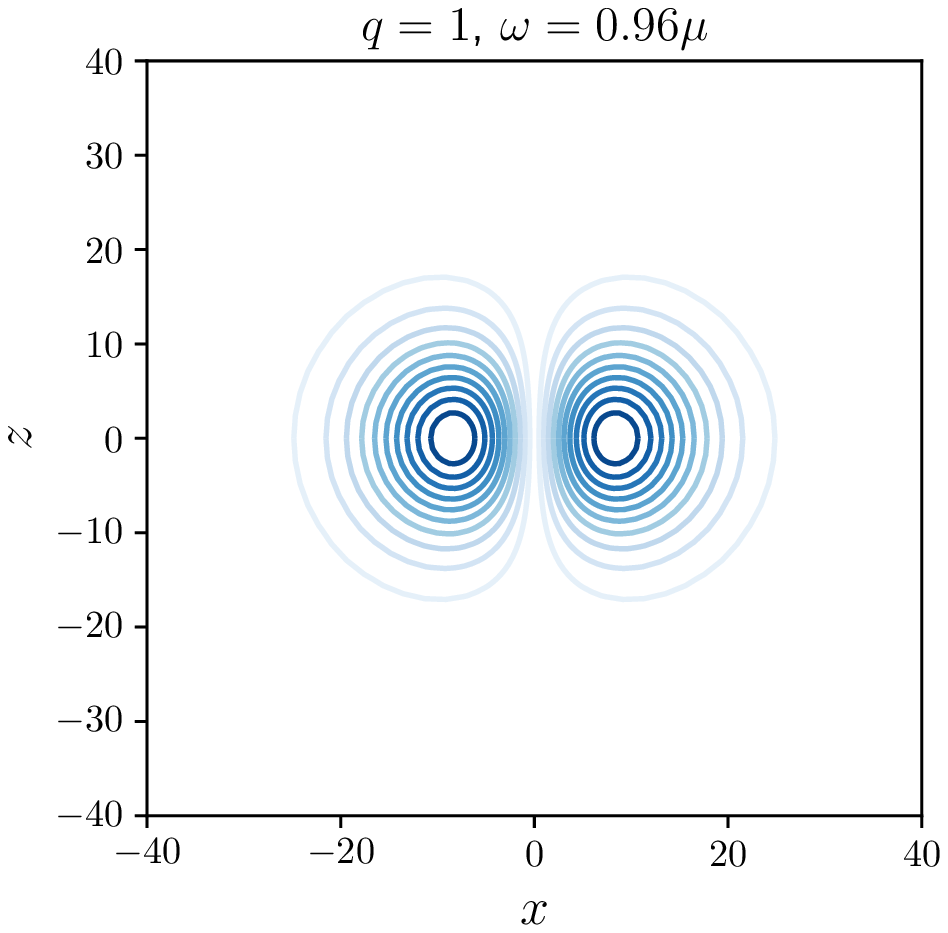} \includegraphics[width=0.3\textwidth]{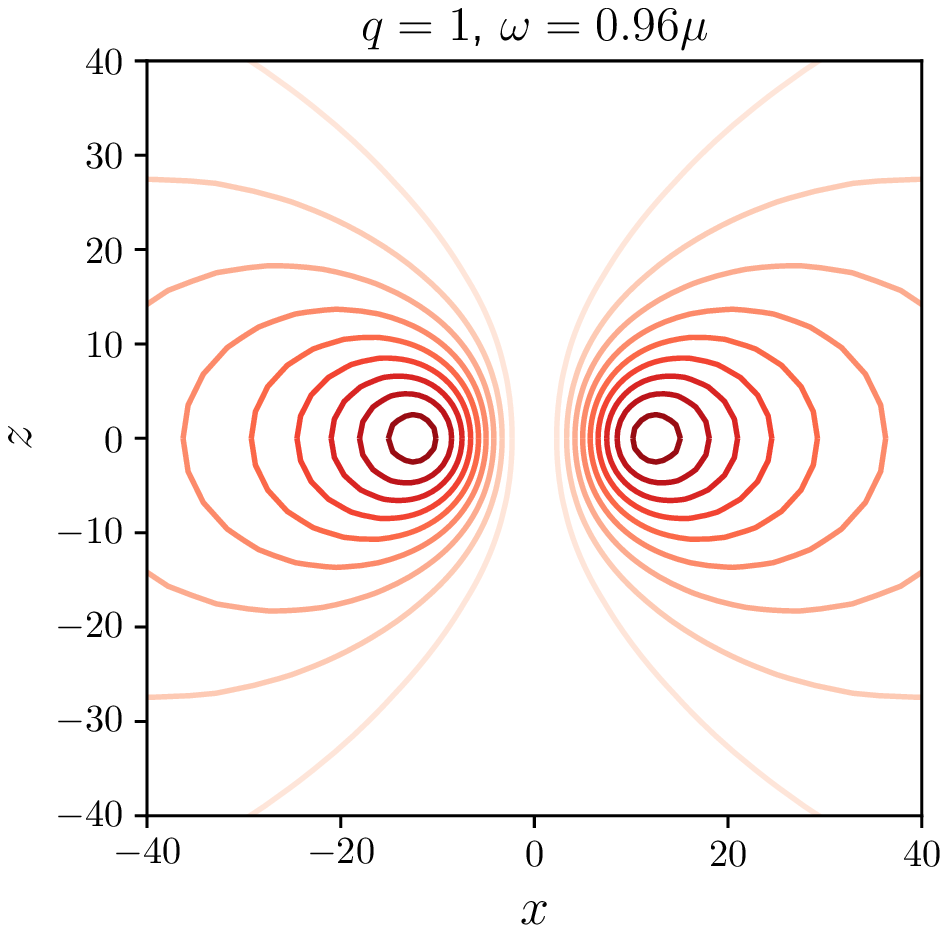}\\
  \includegraphics[width=0.3\textwidth]{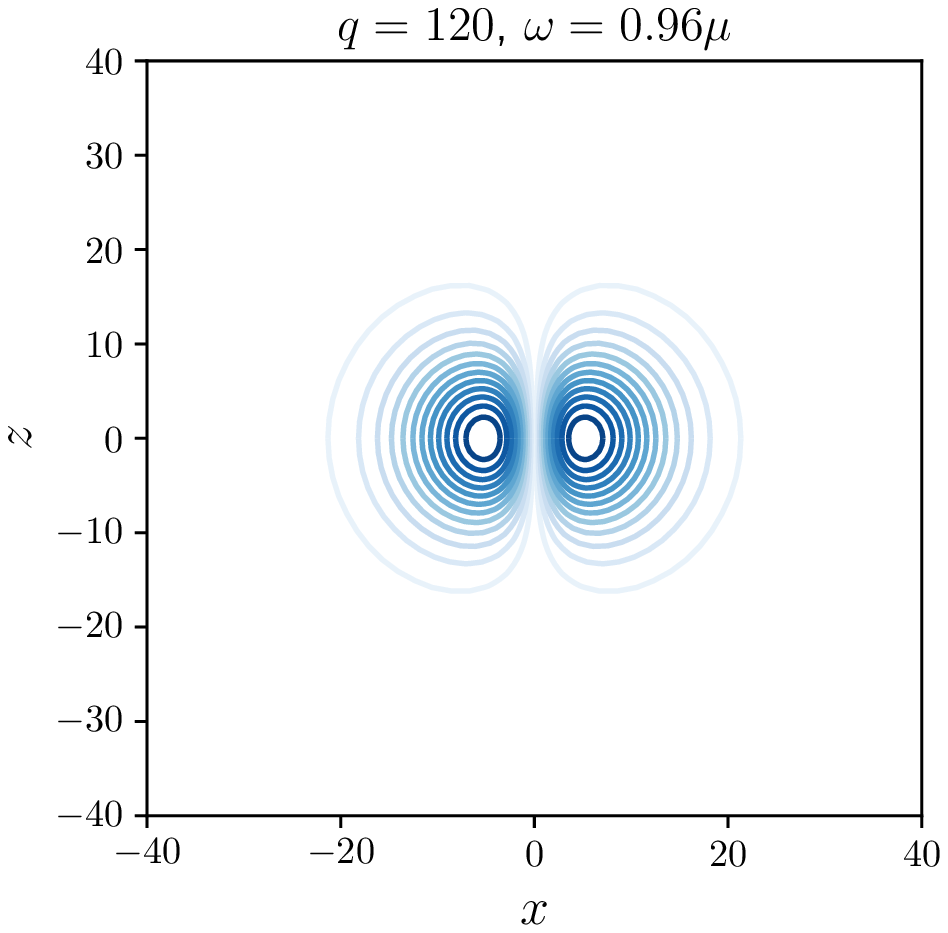} \includegraphics[width=0.3\textwidth]{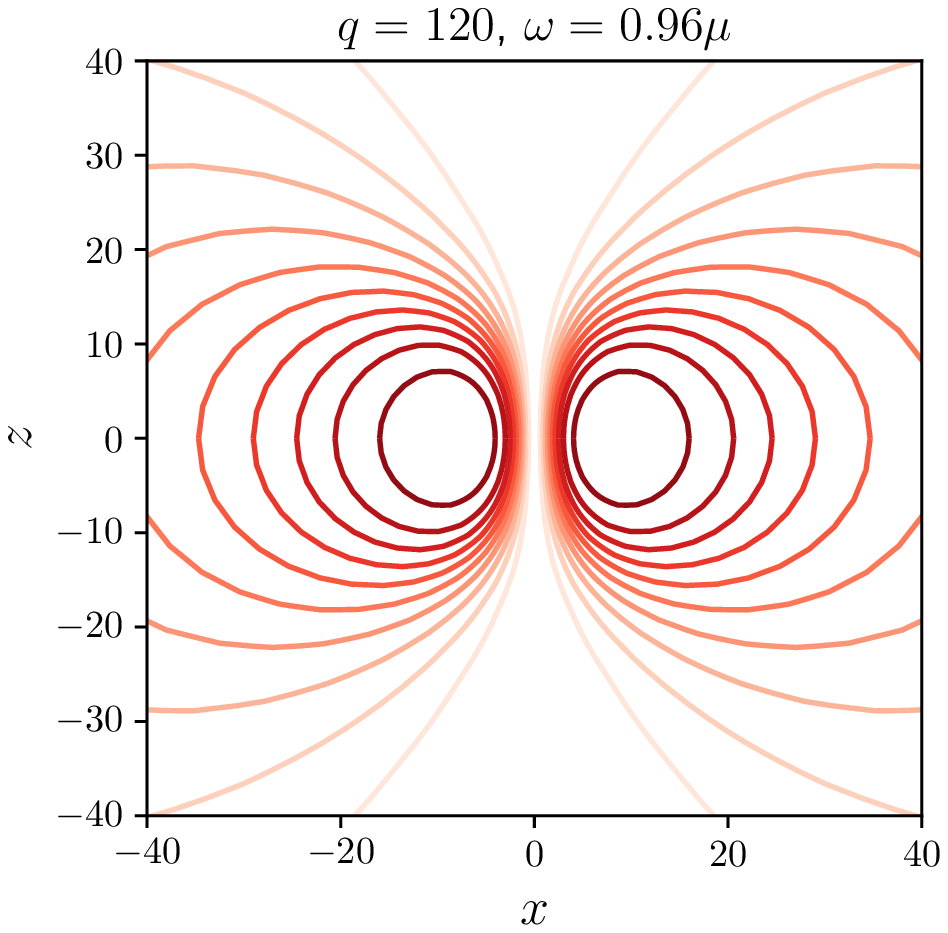}
  \caption{Scalar field $\phi$ isocontours (left) and magnetic field lines (right)
    in a plane of constant $\varphi$ with $x=r\sin\theta$ and $z=r\cos\theta$
    for $m=1$ and $\omega=0.96\mu$ using different values of $q$.
}
\label{fig:contoursm1} 
\end{figure}
%

\begin{figure}
  \includegraphics[width=0.31\textwidth]{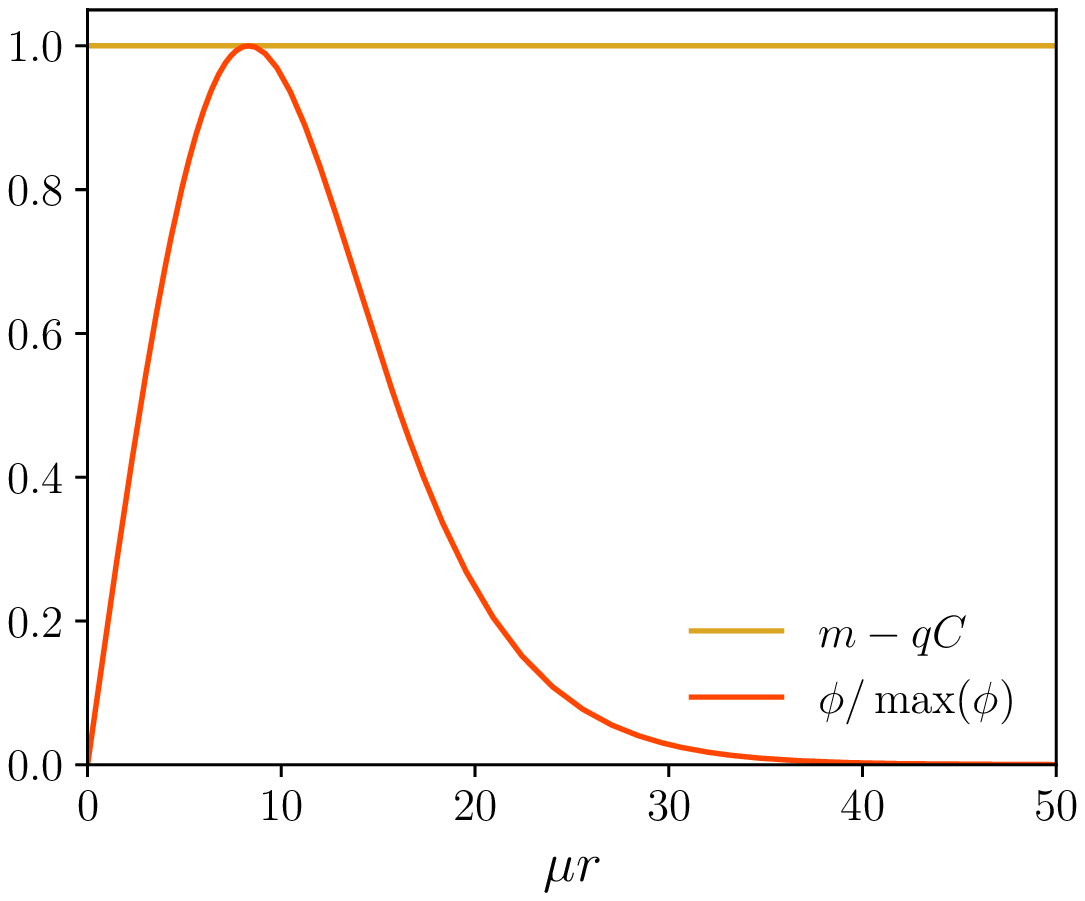} \includegraphics[width=0.31\textwidth]{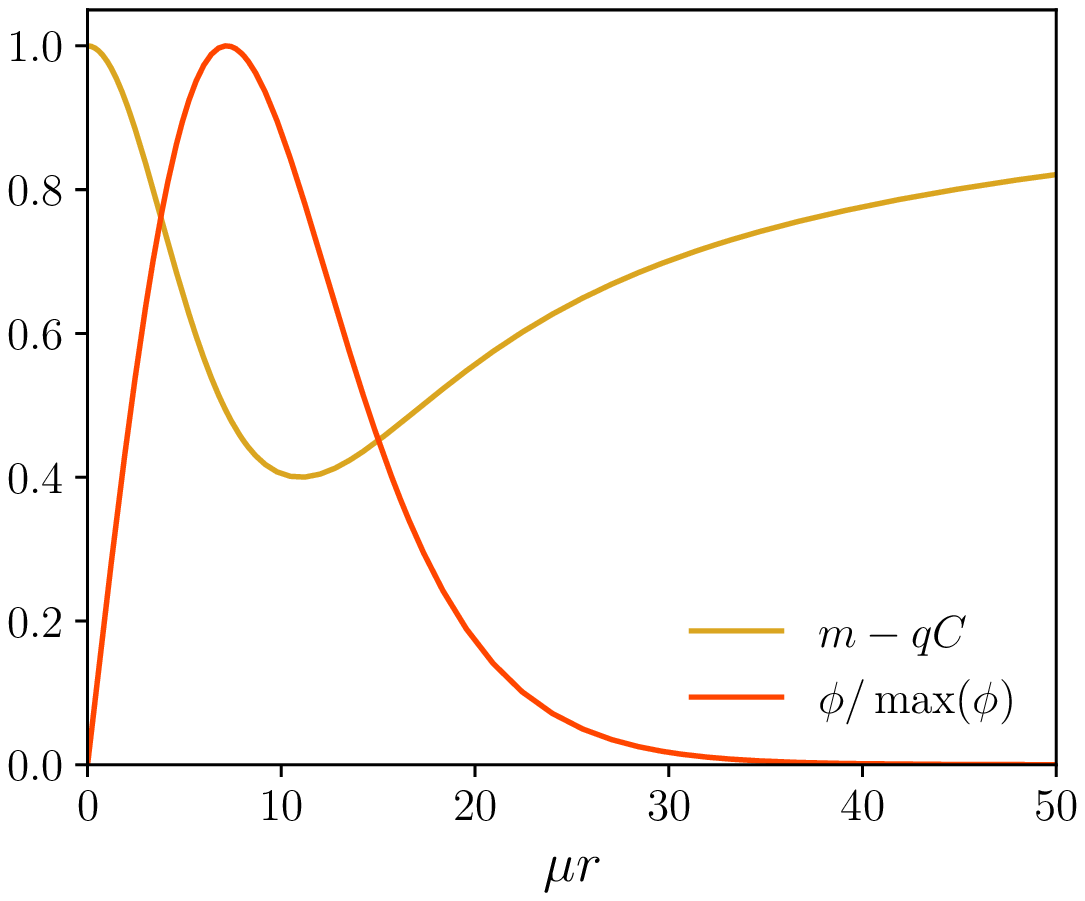}  \includegraphics[width=0.31\textwidth]{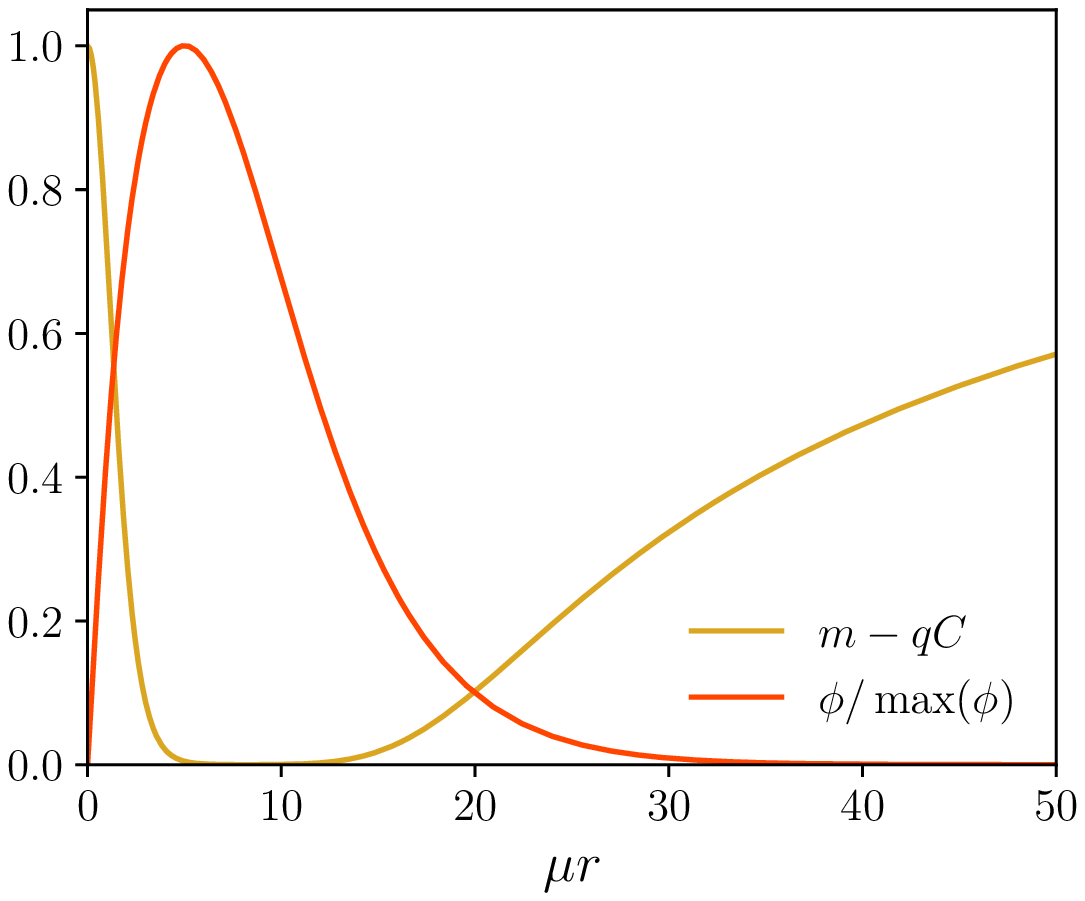}
\caption{Profiles of $\phi$ and $C$ at the equatorial plane for solutions with $m=1$ and $\omega=0.96\mu$. Left panel: Neutral ($q=0$). Center panel: $q=25$. Right panel: $q=140$.
}
\label{fig:perfiles} 
\end{figure}

Typical solutions for magnetized ($q\neq0$) boson stars are presented 
in Fig.~\ref{fig:contoursm1}, where isocountours of the scalar field
function $\phi$ and the $\varphi$ component of $A_\mu$ are plotted
for $m=1$. In the first place we can notice from the $\phi$ contours,
that the star has a toroidal structure just like the $q=0$ case, secondly
we observe from the isocontours of $C$, which can be
interpreted as the magnetic
field lines (see Sec.~\ref{Sec:magnetic}), that the expected poloidal 
magnetic field distribution preserves as $q$ increases, however,
as we will show next in this paragraph, near to
the location of maximum $\phi$ a region of
constant $C$ (zero magnetic field) is formed which grows in size.
This effects on the structure of the star and the 
morphology of the magnetic field can be seen more clearly,
plotting the profiles of the scalar
field and the electromagnetic potential
on the equatorial plane. This is done in Fig.~\ref{fig:perfiles} where
we have plotted $m-qC$ instead of $C$, to observe an interesting property:
Above certain value of $q$ (which in the $\omega=0.96\mu$ case of Fig.~\ref{fig:perfiles} is at 
$q\approx50$, between the second and third panel), the quantity $qC$ approaches
but never exceeds $m$ in a region that grows as $q$ increases. In  
all of the solutions presented in this paper
we have obtained $C(r)<m/q$ for all $r$.

The sources of gravitational field are also enlightening regarding the structure of the
star as well as its global properties, as we will see in the next section. Restricting
to the equatorial plane, $\theta=\pi/2$, the complete
contributions of the energy momentum tensor (see Appendix.~\ref{Sec:3+1}) are given by,

\begin{eqnarray}
E|_{\theta=\pi/2} & = & \left[\frac{\omega^2}{e^{2F_0}}+\frac{(m-qC)^2}{e^{2F_2}r^2}\right]\phi^2+\frac{1}{e^{2F_1}}\left(\frac{\partial \phi}{\partial r}\right)^2+\mu^2\phi^2+E_B,\\
S^r_{\ \, r}|_{\theta=\pi/2} & = & 
    \left[ \frac{\omega^2}{e^{2F_0}} - \frac{(m-qC)^2}{e^{2F_2} r^2} \right]
    \phi^2 + \frac{1}{e^{2F_1}} \left(\frac{\partial \phi}{\partial r}\right)^2 - \mu^2\phi^2+E_B,\\
S^\theta_{\ \, \theta}|_{\theta=\pi/2} & = & 
    \left[ \frac{\omega^2}{e^{2F_0}} - \frac{(m-qC)^2}{e^{2F_2} r^2 } \right]
    \phi^2 - \frac{1}{e^{2F_1}} \left(\frac{\partial \phi}{\partial r}\right)^2 - \mu^2\phi^2 - E_B\\
S^\varphi_{\ \, \varphi}|_{\theta=\pi/2} & = & 
    \left[ \frac{\omega^2}{e^{2F_0}} + \frac{(m-qC)^2}{e^{2F_2} r^2} \right]
    \phi^2 - \frac{1}{e^{2F_1}} \left(\frac{\partial \phi}{\partial r}\right)^2 - \mu^2\phi^2 + E_B 
\end{eqnarray}

Where we have defined $E_B$ as the purely electromagnetic 
contribution to the energy density (at the equatorial plane),

\begin{equation}
E_B:=\frac{1}{2r^2e^{2F_1+2F_2}}\left(\frac{\partial C}{\partial r}\right)^2.
\end{equation}

\begin{figure}
  \includegraphics[width=0.325\textwidth]{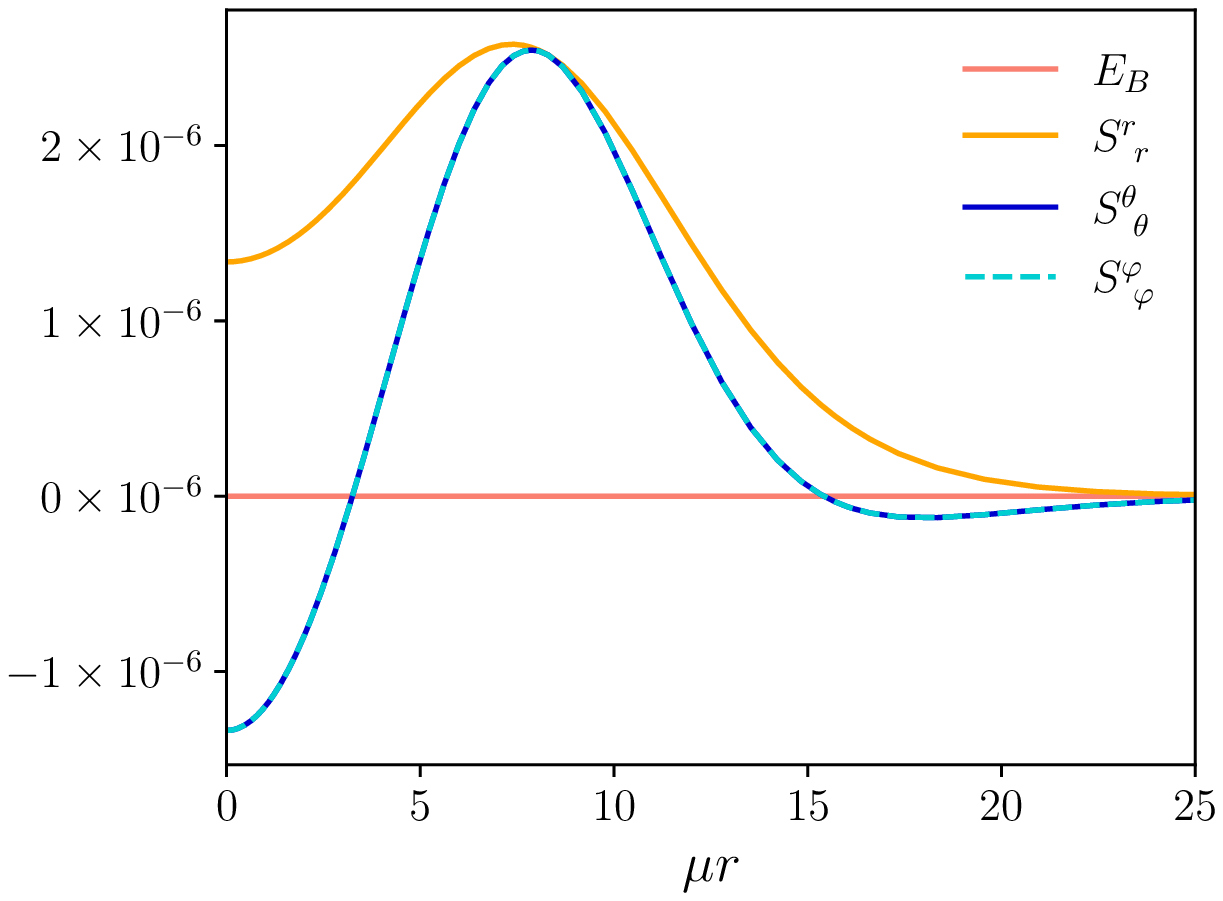} \includegraphics[width=0.325\textwidth]{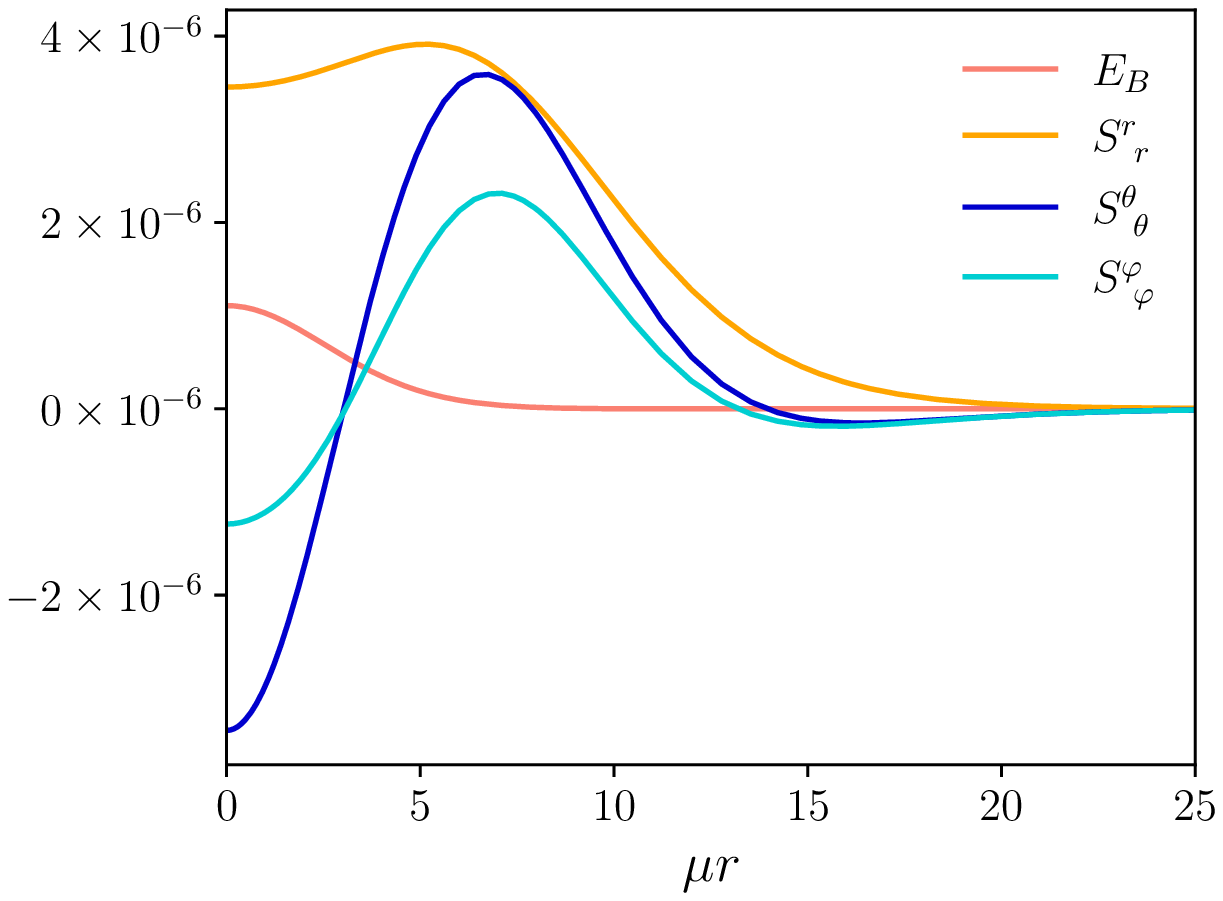}  \includegraphics[width=0.325\textwidth]{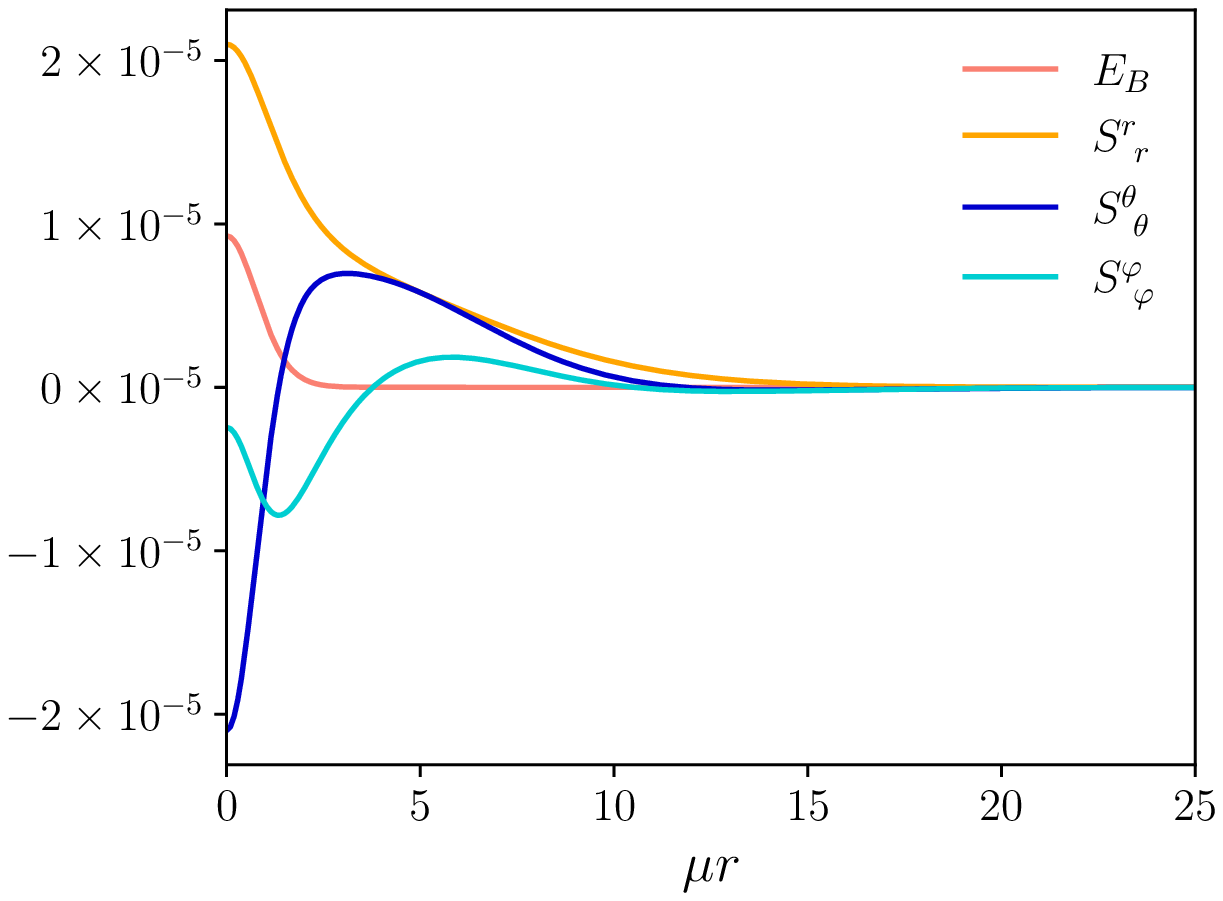}
\caption{Source terms at the equatorial plane for boson stars with $m=1$ and $\omega=0.96\mu$. Left panel: Neutral ($q=0$). Center panel: $q=25$. Right panel: $q=140$.
}
\label{fig:fuentes} 
\end{figure}

Regarding the stress tensor components,
which are usually identified as components of the pressure,
the system is completely anisotropic even in the $q=0$ case, 
where for instance the difference
$S^\theta_{\ \, \theta} - S^\varphi_{\ \, \varphi} \propto \phi^2e^{-2 F_2}/r^2$ is not zero, 
although suppressed by $r^2$. In the left panel of Fig.~\ref{fig:fuentes} we plot
the sources of the $q=0$ case where in fact the difference between $S^\theta_{\ \, \theta} - S^\varphi_{\ \, \varphi}$
cannot be appreciated. The middle and right panels of Fig.~\ref{fig:fuentes} show the magnetized
$q=25$ and $q=140$ cases respectively. Notice that the extrema of $S^\varphi_{\ \, \varphi}$ 
decrease in magnitude with respect to the extrema of $S^\theta_{\ \, \theta}$ and $S^r_{\ \, r} $.
In particular near $r=0$, the minimum of $S^\varphi_{\ \, \varphi}$ increases 
due to the $E_B$ contribution. The behavior of the pressure term $S^\varphi_{\ \, \varphi}$ 
is relevant in the analysis of the effect of $q$ on  global quantities, as for example the magnetic
dipole moment and the total mass. This will be discussed in the next section.

\subsection{Sequence of magnetic boson stars}

For $m=1,2$ we obtained a family of configurations by means of slowly varying
the parameters of the solution starting from a Newtonian solution, as 
stated before. First, we have verified that in the case $q=0$, $m=1$ we obtain 
the known sequence of toroidal static boson stars \cite{Sanchis-Gual:2021edp}. 
Thereafter,
starting from this set of solutions, we have slowly increased 
the value of $q$, generating in this way sequences of magnetized boson stars.

\begin{figure}
  \includegraphics[width=0.4\textwidth]{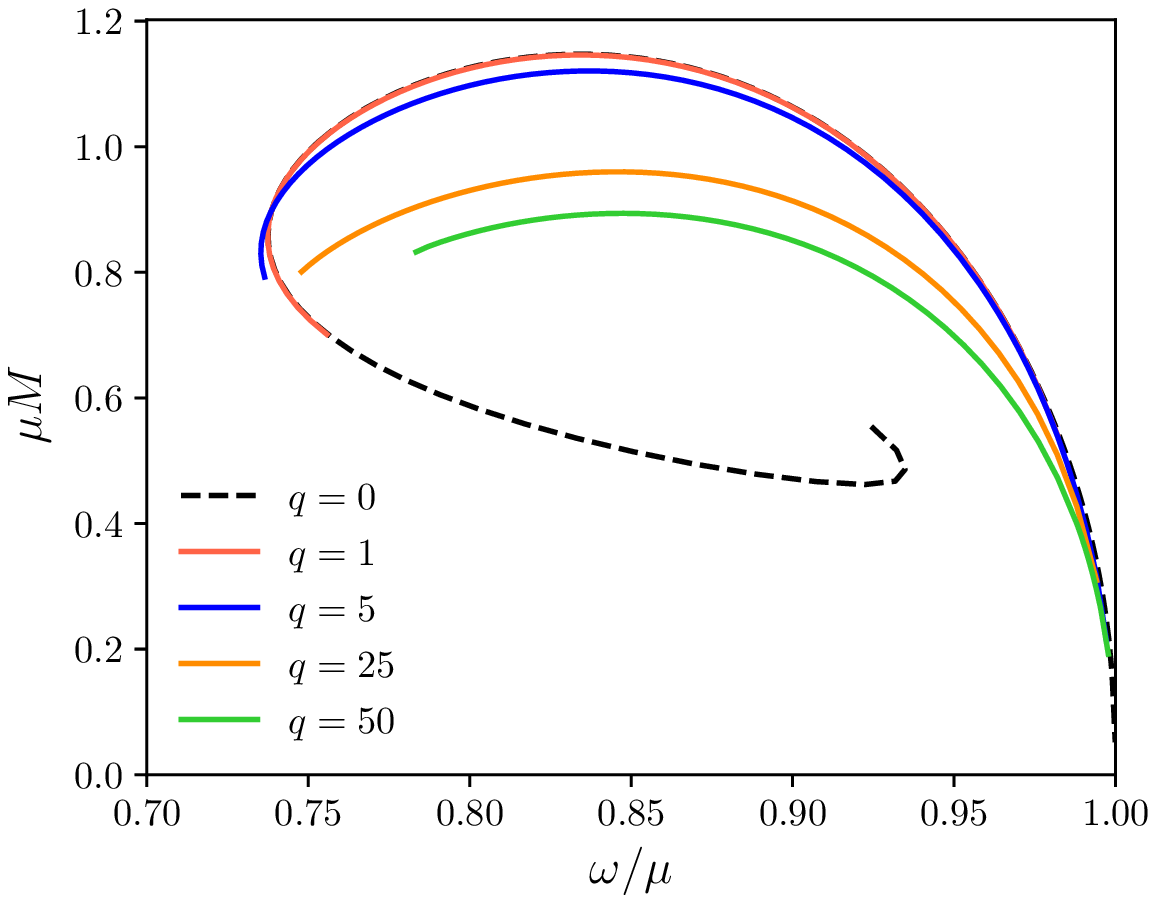}\includegraphics[width=0.415\textwidth]{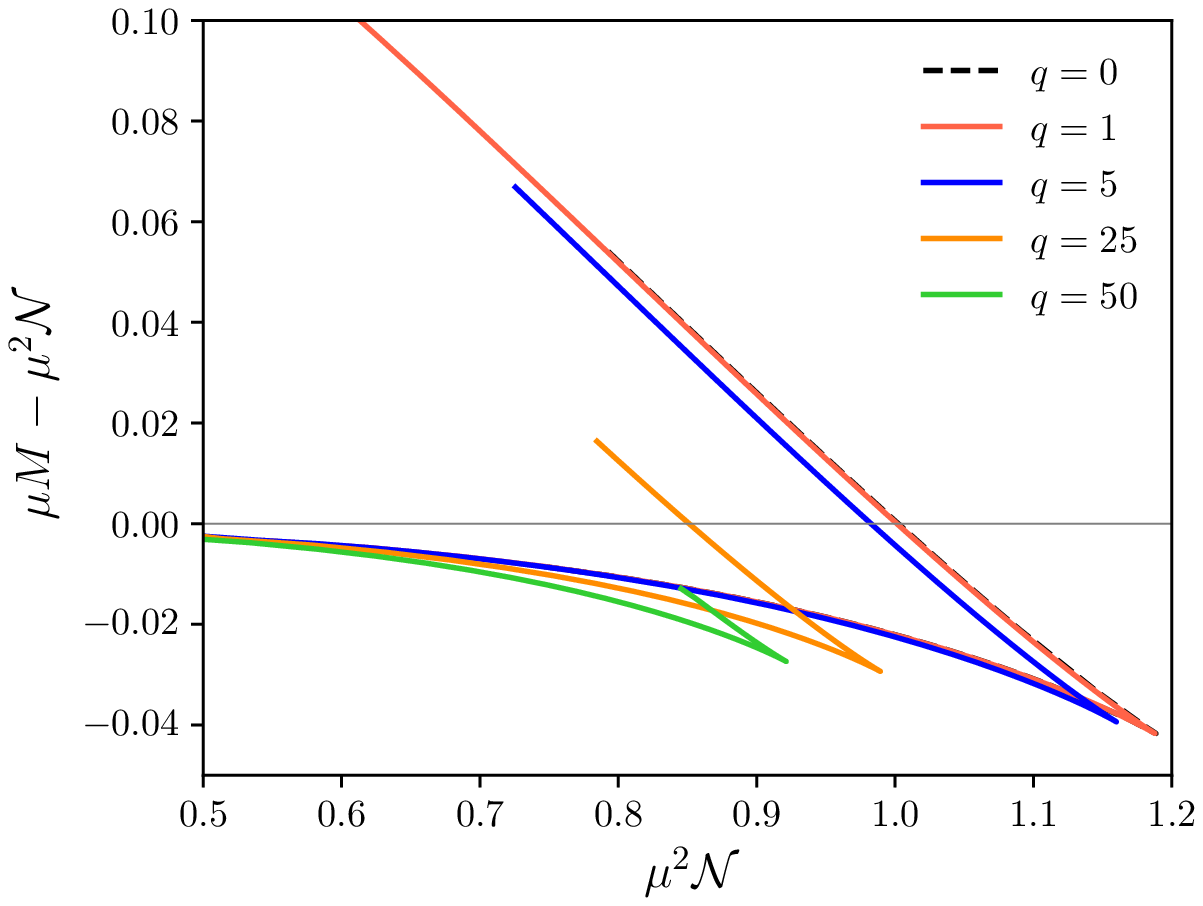}
  \caption{Left panel: Frequency \textit{vs}. mass for the $m=1$ magnetostatic boson star solutions
    using different values of the coupling constant $q$.
    Right panel: Binding energy $M-\mu N$ for $m=1$.
}
\label{fig:omegavsM} 
\end{figure}

In Fig.~\ref{fig:omegavsM}, the global quantity $M$ is shown \textit{vs}.
the scalar field frequency $\omega$ for $m=1$ and five chosen values for $q$.
Some interesting aspects arise from these solutions:
firstly the mass of the star decreases monotonically with 
$q$; this is the opposite behavior to that obtained
in models of neutron stars with magnetic fields
\cite{Bocquet:1995je,Cardall:2000bs}, where
their structure begins from spherical morphology at zero magnetic field (for the static case),
and flattens, increasing the circumferential radius of the star, as the magnitude of the 
magnetic fields increase, with a corresponding increase in the mass of the star.
The observed structure dependence on $q$ of the magnetized boson stars,
is also opposite to the corresponding dependence of charged
boson stars as discussed in the previous section. However,
such observed decrease on the size and total 
mass $M$ as the coupling constant $q$ grows, is also observed in
the magnetized Bose-Einstein condensate stars~\cite{QuinteroAngulo:2018sem}.

The second aspect to consider about Fig.~\ref{fig:omegavsM}, regards the existence of
equilibrium configurations with coupling constant  $q$ above the value
$q_\mathrm{crit}\approx 1/\sqrt{2}$. Indeed, as  reported in
\cite{Pugliese:2013gsa,Collodel:2019ohy}, that value was an upper limit
for stable charged and stable rotating-charged boson stars.
In our model without total charge, that limit is overcome. 
This interesting result is related to the fact that the Lorentz force,
\begin{equation}
f_\nu:=F_{\nu\alpha} J^\alpha=-\nabla_\mu (T^{\mathrm{EM}})^{\mu}_{\ \, \nu},
\end{equation}
points everywhere outwards for the charged mini-boson stars, while for the 
magnetostatic boson star it only points outward near the origin and points inward
outside the main distribution of scalar field. Therefore, the nonrelativistic
argument regarding Coulomb repulsion \textit{vs.} gravitational attraction
does not apply here. Instead, it is the stress anisotropy that ultimately
determines the structure and global properties of the star, as we will
see below in relation to the decrease in the total mass. Numerically
we have not obtained any limiting value for the parameter $q$, the
equations are difficult to solve for large values of the coupling
constant due to the resolution required at the ``edges of the
plateau'' that forms in the function $C$ (see the right panel of Fig.~\ref{fig:perfiles}).

\begin{figure}
  \includegraphics[height=0.225\textheight]{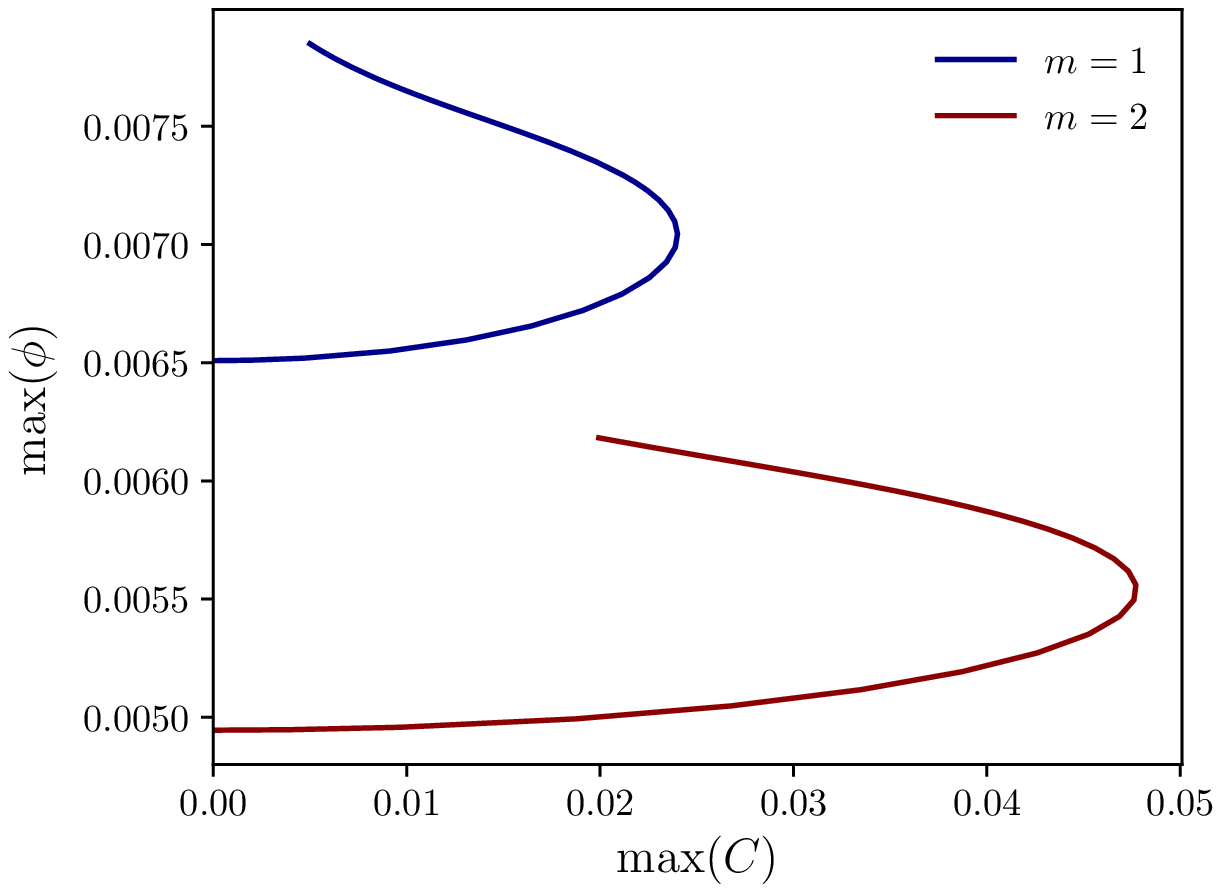}\includegraphics[height=0.225\textheight]{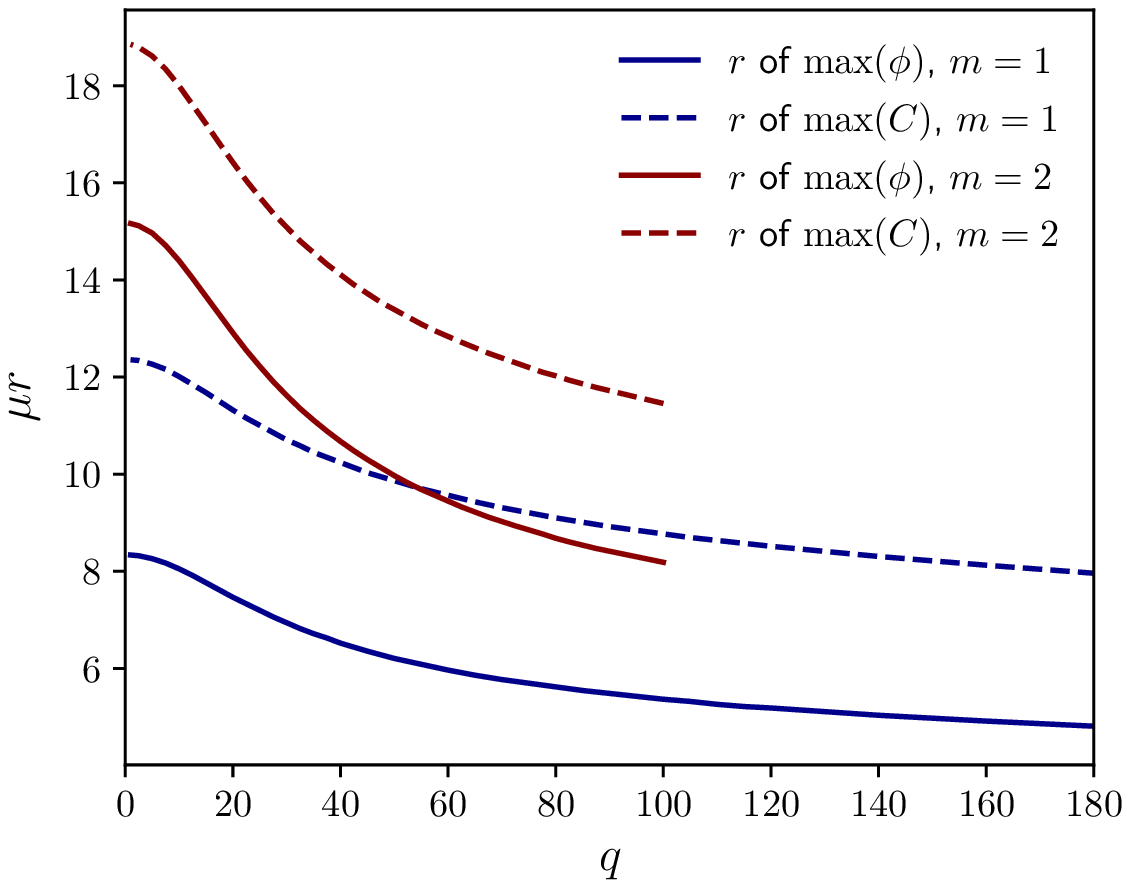}
  \caption{Sequence of solutions constant $\omega$ and increasing coupling $q$.
    Given $q$ we locate in the solution
    for the maximum of $\phi$ and $C$ and its location.
    Left panel: Maximum value of the scalar field with respect to the maximum
    value of the four potential function $C$ for sequence of solutions with
    $m=1$ and $m=2$ with $\omega=0.96\mu$. 
    Right panel: Radius at which the maxima of $\phi$ and $C$ is attained, as a function 
    of the coupling constant $q$.
}
\label{fig:maxvsmax} 
\end{figure}

Anisotropic pressures are essential to obtain equilibrium 
configurations with high compactness and large values for the mass. 
In \cite{Alcubierre:2021psa} (see \cite{Raposo:2018rjn} for recent discussion on
fluid anisotropic stars and \cite{Andreasson:2007ck} for shell-type configurations in the
Einstein-Vlasov system)
it was shown that for $\ell$-boson stars, small radial pressures
and big tangential pressures are related to an increase in the
mass and radius of the star in a way that resembles the forces on an arch.
In our case, to understand the decrease
in size of the magnetic boson stars,
we start by noticing that the tangential pressure
is composed by two different
contributions, $S^\theta_{\ \, \theta}$ and $S^\varphi_{\ \, \varphi}$, which act
together with the radial pressure (and with the Lorentz force in some region)
against gravity in order to support the configuration. The toroidal shape, which
is made possible by the $S^\varphi_{\ \, \varphi}$ contribution, shrinks with increasing
$q$ - right panel Fig.~\ref{fig:maxvsmax}, given that the electromagnetic
contribution to the energy-momentum tensor
in the region where the scalar field concentrates, 
is bigger for $S^r_{\ \, r}$ than for the tangential components and in
particular with respect to $S^\varphi_{\ \, \varphi}$,
as discussed in Sec.~\ref{Sec:structure}.
Our results indicate that this reduction in the size of the star
is accompanied by a reduction in the total mass that the boson star
can support.

\begin{figure}
  \includegraphics[width=0.4\textwidth]{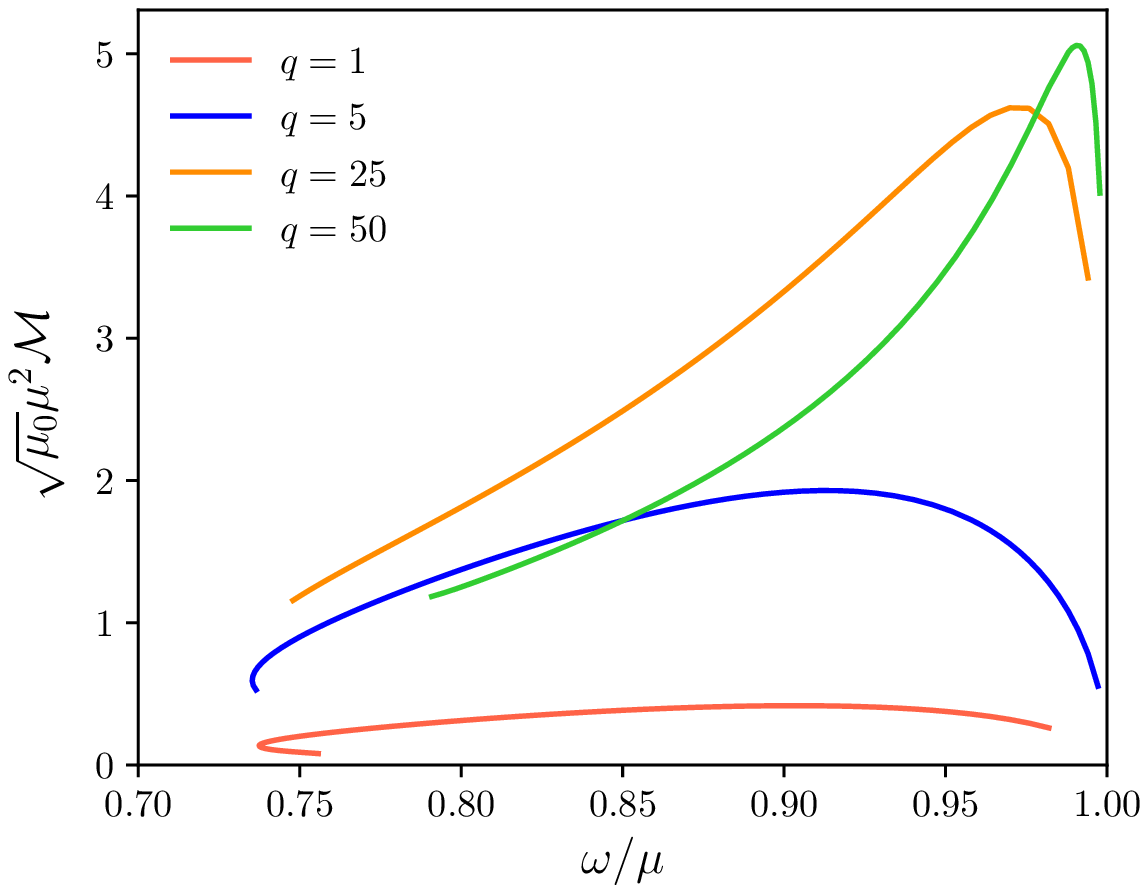} \includegraphics[width=0.4\textwidth]{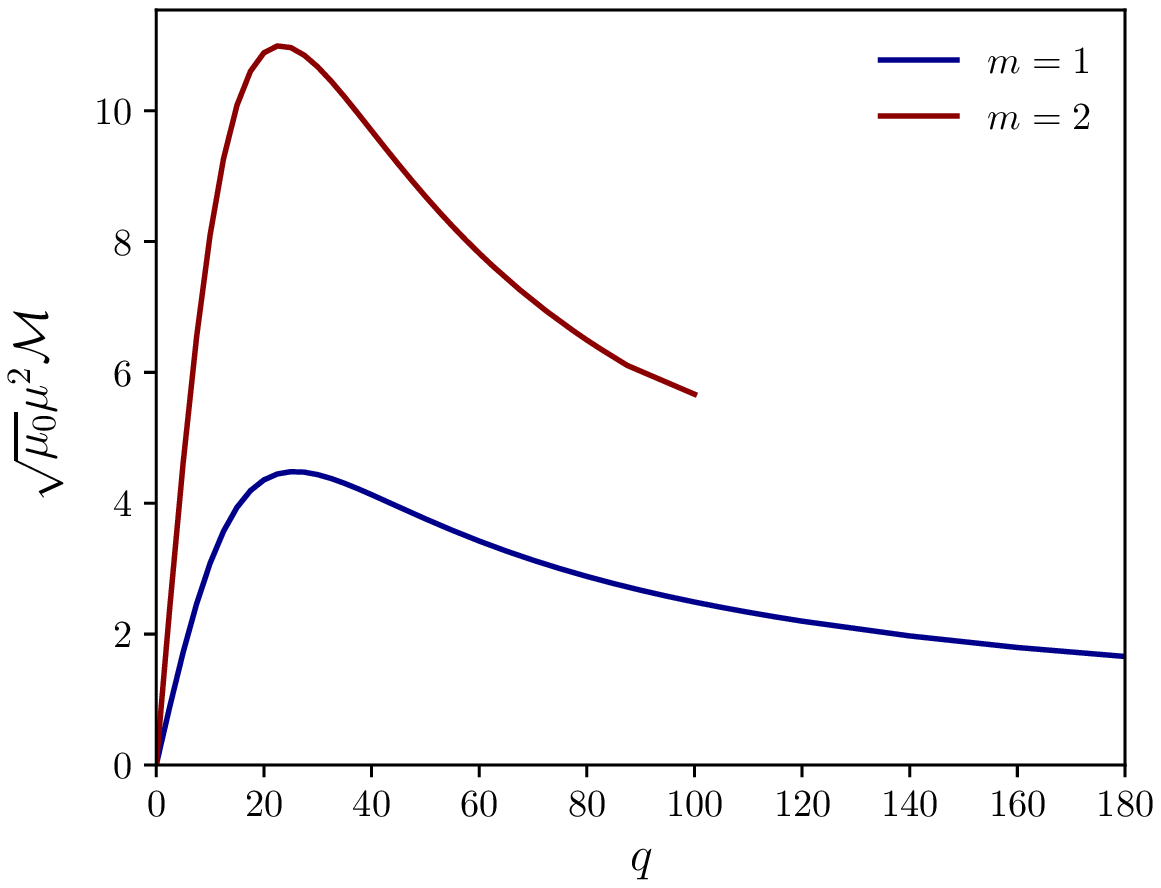}
  \caption{Left panel: Frequency vs. magnetic dipole moment $\mathcal{M}$ for the
    $m=1$ magnetostatic boson star solutions
    using different values of the coupling constant $q$.
    Right panel: Magnetic dipole moment $\mathcal{M}$ as a function of $q$ for configurations
    with $m=1$ and $m=2$ and a fixed value of $\omega$.
}
\label{fig:omegavsb} 
\end{figure}

Fig.~\ref{fig:maxvsmax} shows properties
of the scalar field, $\phi$ and of the electromagnetic one, $C$ \textit{vs.} $q$.
More precisely, the figure shows the maximum of the functions
and the coordinate at which the maximum is attained. From the right panel
we appreciate the decrease in size of the torus as a function of $q$ for
fixed $\omega$.
In the left panel we see another important property of the
solutions: $\max(C)$ reaches a maximum value and then begins to decrease
with $q$. As a consequence, the magnetic dipole moment $\mathcal{M}$
which can be obtained from the asymptotic behavior of the electromagnetic
potential $A_\mu$,
\begin{equation}
  A_\mu dx^\mu\sim \frac{\mu_0}{4\pi}\frac{\mathcal{M}\sin^2\theta}{r}d\varphi,
\end{equation}
reaches a maximum and decreases thereafter.
This behavior is shown in the right panel of
Fig.~\ref{fig:omegavsb} for the sequence of $m=1$
solutions using four selected values of $q$ and
in the same panel can be seen
for fixed $\omega$, $m=1,2$  and several values of $q$.
%
%
One can note from these
plots that larger values for the maximum of $\mathcal{M}$ are obtained,
closer to $\omega=\mu$, as $q$ increases. In Table \ref{maxMM} we provide data
of the maximum $\mathcal{M}$ configuration for a sample of values for
the coupling constant.
For the explored $q\gg1$ configurations, the maximum of $\mathcal{M}$ 
increases more slowly, for example, the $q=200$ case, with $\max(\mathcal{M})=4.84$,
is not far from the value obtained 
for the $q=50$ case, plotted in Fig.~\ref{fig:omegavsb}, so the maximum dipole moment
seems to tend asymptotically to a finite value, 
however to establish with precision the limit, it would be necessary to solve
the equations in the limiting case $q\rightarrow\infty$, which is beyond the 
scope of this paper.

\begin{table}\centering
  \caption{Maximum mass and magnetic dipole configurations. \label{maxMM}}
  \begin{tabular}{l c c c c c c c}
    \hline
    \hline
    $\quad \qquad$& $\quad q\quad$     & $\quad\sqrt{\mu_0}\mu^2\mathcal{M}\quad$     &  $\quad\omega/\mu\quad$      & $\quad\mu M\quad$      & $\quad\mu^2 \mathcal{N}\quad$  &  $\quad\max(\phi)\quad$  &  $\quad\mu r_{\max(\phi)}\quad$\\
    \hline
    Maximum $\mathcal{M}$ \\
    \hline
    & $0.1$  &  0.0418  &  0.905  &  1.048  &  1.077  &  0.0156  &  4.66\\
    & $1$   &  0.417  &  0.905  &  1.047  &  1.076  &  0.0156  &  4.66 \\
    & $5$  &  1.93  &  0.911  &  1.015  &  1.041  &  0.0147  &  4.82  \\
    & $25$   &  4.62  &  0.970  &  0.627  &  0.633  &  0.00492  &  9.08 \\
    & $50$   &  5.06  &  0.990  &  0.363  &  0.363  &  0.00154  &16.8\\
    & $100$  &  5.17  &  0.997  &  0.195  &  0.195 &  0.000454  &  30.1 \\
    & $200$  &  4.84  &  0.999  &  0.084  &  0.085  &  0.000132  &  36.0\\
    \hline
    Maximum $M$\\
    \hline
    & $0$  &  0  &  0.840  &  1.147  &  1.189  &  0.0290  &  2.76\\
    & $0.1$  &  0.0368  &  0.834  &  1.147  &  1.189  &  0.0303  &  2.65\\
    & $1$  &  0.366  &  0.834  &  1.146  &  1.188  &   0.0303  &  2.65 \\
    & $5$  &  1.66  &  0.839  &  1.121  &  1.160  &  0.0296  &  2.69  \\
    & $25$  &  2.46  &  0.848  &  0.960  &  0.989  &  0.0312  &  2.22\\
    & $50$  &  1.69  &  0.848  &  0.894  &   0.921  &  0.0337  &  1.85\\
    \hline
    \hline
  \end{tabular}
\end{table}

Finally, we wonder about the possibility of determining
the magnetic dipole moment
from the asymptotic behavior of the metric functions. For example, in the charged
mini-boson star, the total electric charge of the configuration can be read off the
$g_{rr}$ component by comparing with the Reissner-Nordstr\"om solution \cite{Jetzer:1989av}.
Some electrovacuum exact solutions (in General Relativity) for a mass 
endowed with a magnetic dipole moment have
been obtained in the literature, as for example the Gutsunaev-Manko
\cite{GUTSUNAEV1987215} and Bonnor \cite{Bonnor1966AnES} two-parameter family of solutions.
However, analyzing our solutions, we obtain that they do not match with neither of those metrics, 
for instance the lapse function in all of the solutions that we generate
has the asymptotic behavior 
$N^2=1-2M/r+\alpha/r^2+\mathcal{O}(1/r^3)$, with $\alpha$ some constant,
while according to \cite{GUTSUNAEV1987215}, the lapse of the Gutsunaev-Manko metric
goes as $N^2=1-2M/r+\mathcal{O}(1/r^3)$. On the other hand, comparison with the Bonnor 
solution, for which $N^2=1-2M/r+M^2/r^2$, could seem to be a better alternative,
however we obtain from the analysis of our solutions that the coefficient $\alpha$
is a function of both of $M$ and $\mathcal{M}$ and in the $\mathcal{M}=0$ case,
it is not proportional to $M^2$. Therefore the magnetic dipole moment of
the star cannot be obtained from the the metric components by comparison with any of the
mentioned exact solutions, allowing us to conclude that our solutions differ from those two spacetimes.


\section{Magnetic field}
\label{Sec:magnetic}

The electric and magnetic field as measured by an observer
whose four-velocity is $n^\mu$ (Eulerian observer) are given by
the formulas $E_\mu=F_{\mu\nu}n^\nu$ and
$B_\mu=-\frac{1}{2}\epsilon_{\mu\nu\alpha\beta}n^\nu F^{\alpha\beta}$.
where $\epsilon$ is the Levi-Civita tensor. For the metric~(\ref{eq:metric})
and the electromagnetic four-potential~(\ref{eq:ansatzA}), we obtain $E_\mu=0$
as expected and
%
\begin{equation}
  B_\mu dx^\mu=\frac{e^{-F_2}}{\sin\theta}\left(\frac{1}{r^2}\frac{\partial C}{\partial\theta}\ dr
  -\frac{\partial C}{\partial r}\ d\theta\right).
\end{equation}

\begin{figure}
    \includegraphics[width=0.33\textwidth]{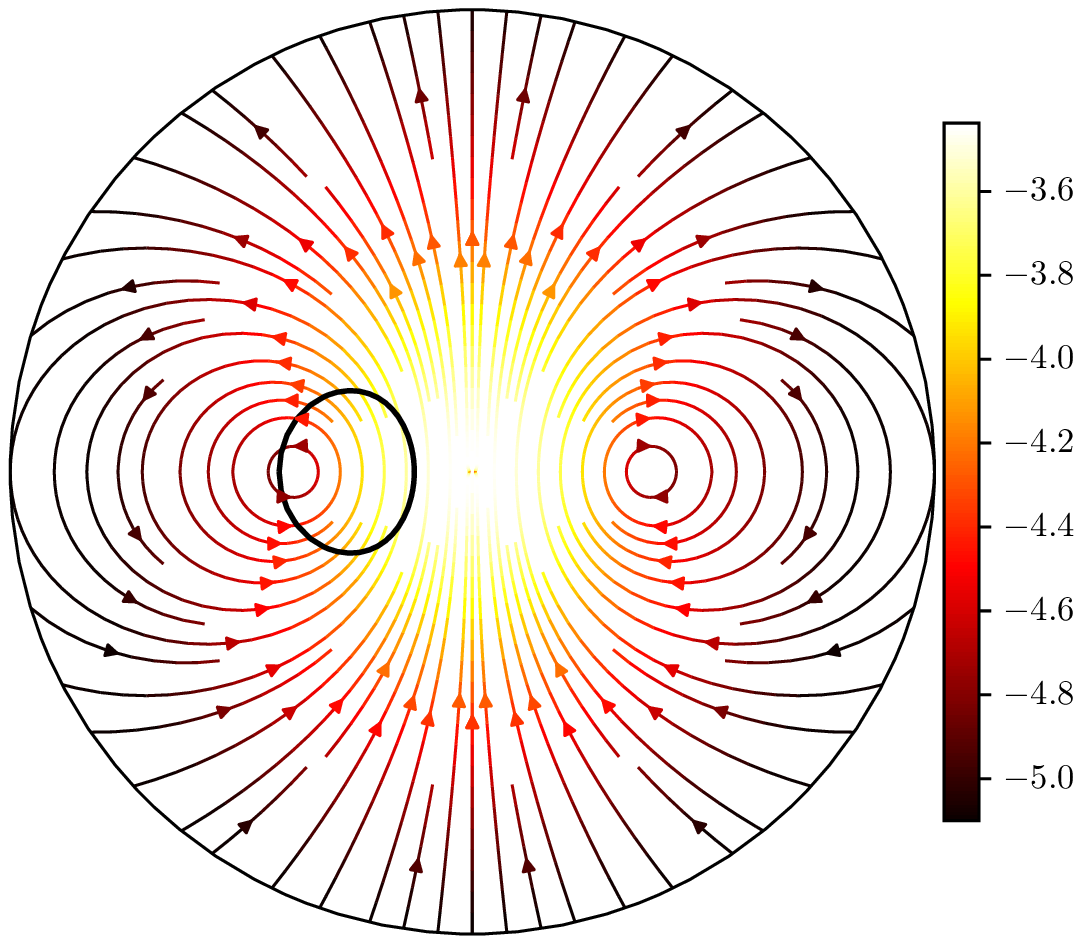}\includegraphics[width=0.33\textwidth]{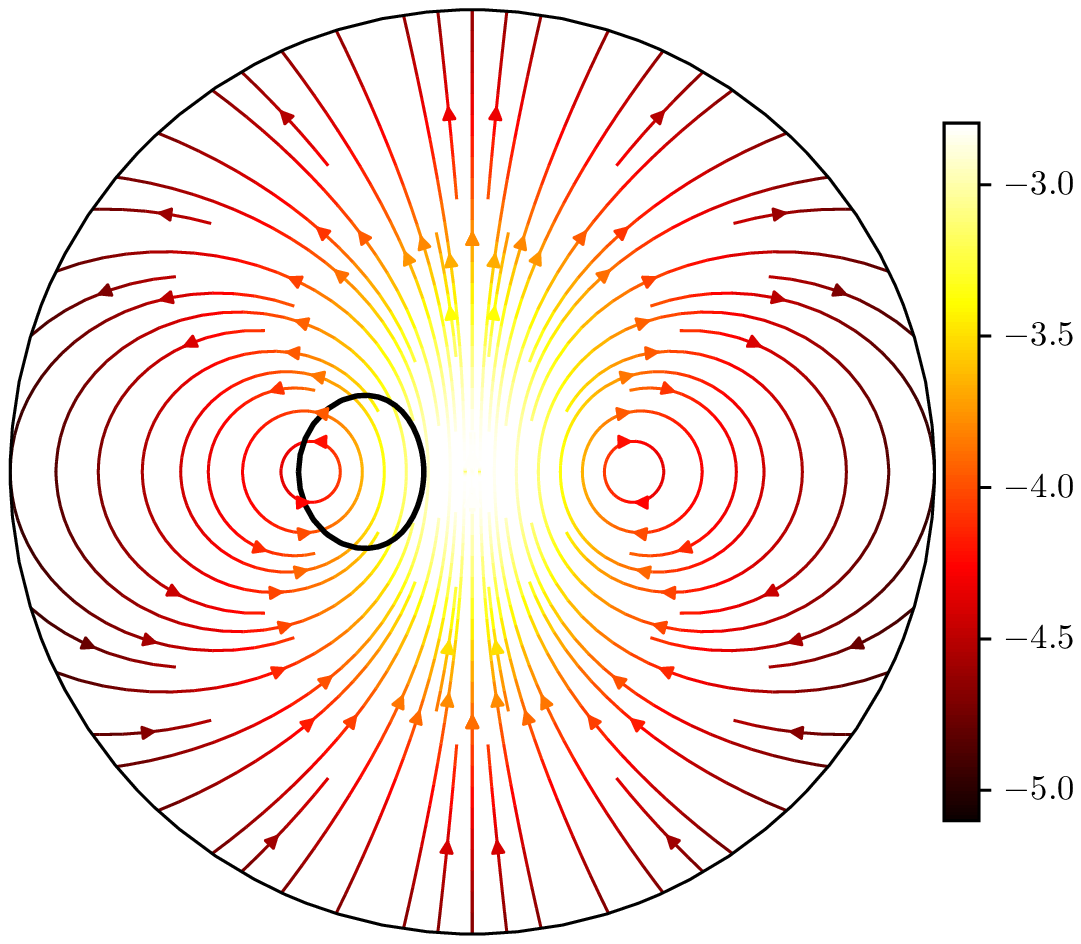}\includegraphics[width=0.33\textwidth]{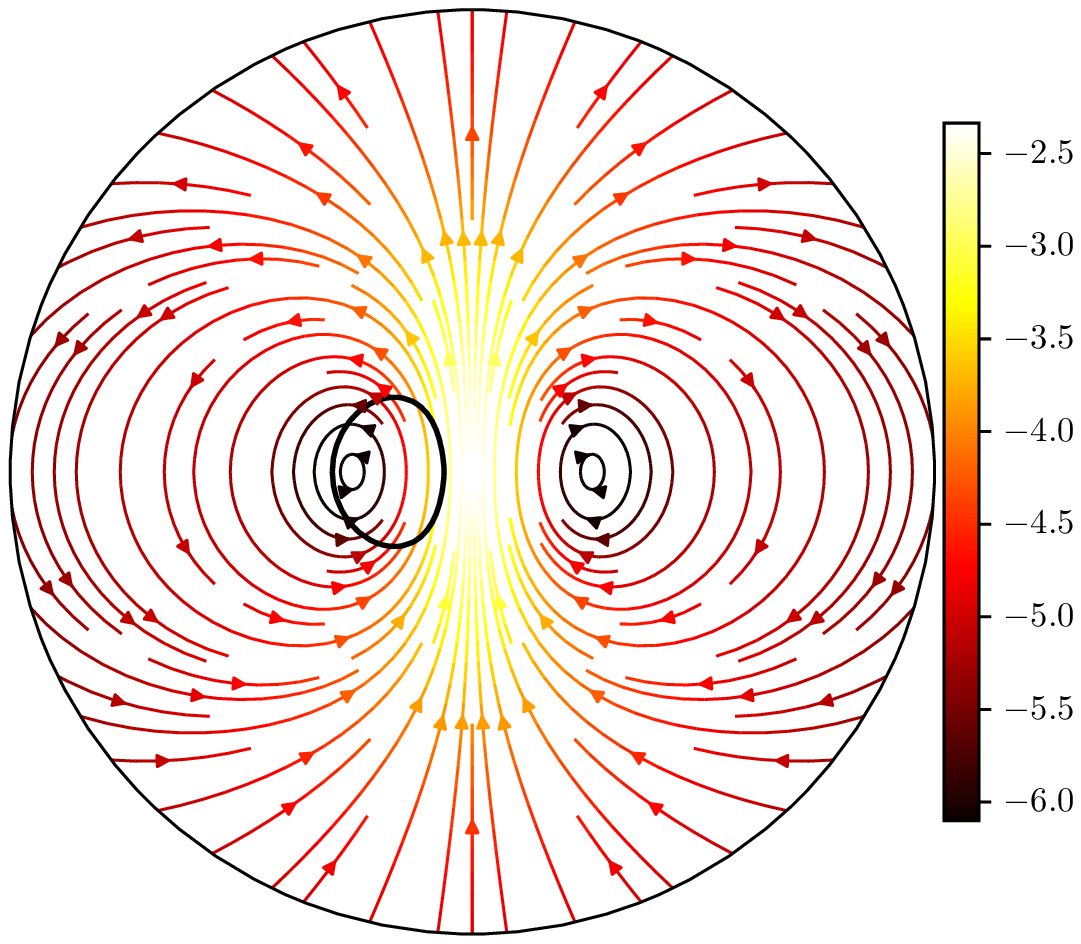}
  \caption{Magnetic field in a plane of constant $\varphi$ for configurations with $m=1$ and $\omega=0.96\mu$, 
  for the coupling constant with values (from left to right) $q=5$, $25$ and $140$. The thick black line 
  correspond to the density isocountour $E=0.5 \max(E)$.
  The color bar indicates the norm of the quantity $\log_{10}[B^i/(\mu\sqrt{\mu_0})]$ 
  The radius of the circle is $\mu r=32$. 
}
\label{fig:B} 
\end{figure}

Some examples of $B^i$ for configurations with $m=1$ are given in Fig.~\ref{fig:B}.
The distribution of the vector field resemble that of the magnetic field around 
a finite size current loop. For reference we also plot the isocontour of half the maximum 
value of the energy density. Rotation of this curve around the $z$ axis generates a torus.
The figure also shows the region of zero magnetic field that forms in 
configurations with high values of $q$, where $m-qC\approx0$,
see for instance the black line region with $\sqrt{B^iB_i}/(\mu\sqrt{\mu_0})<10^{-6}$
inside the torus in the right panel of Fig.~\ref{fig:B}.

We have seen in the previous section that as $q$ get closer to zero, the magnetic moment decreases
and the maximum values of $\phi$ and $C$ are reached at larger radii.
This explains why
some of the configurations with relative low values of $q$,
as for example the $q=5$ and $q=25$ cases (Fig.~\ref{fig:B}, left and central panels),
do not have the maximum 
of $B^i$ at the center of the star but in a toroidal region around the center, while other configurations
as for instance the $q=140$ case (Fig.~\ref{fig:B}, right panel), posses magnetic fields 
concentrated in a central region with maximum along $\theta=0$.

\begin{figure}
  \includegraphics[width=0.4\textwidth]{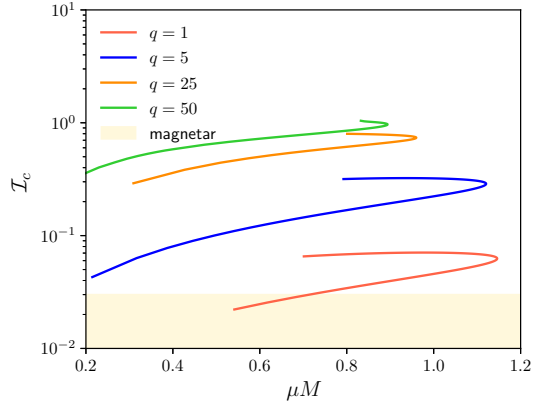}
  \caption{Dimensionless quantity $\mathcal{I}$ defined in Eq.~(\ref{eq:I}),
    evaluated at $r=0$
    for configurations with $m=1$ and different values of $q$.
    The yellow region corresponds to the upper region of the interval $\mathcal{I}$
    at the center of strongly magnetized neutron stars (see the text for details).
}
\label{fig:magB} 
\end{figure}

Until now it has not been required to specify the value
for the mass parameter of the scalar field
given that solutions with different $\mu$ are related to each other
by rescaling rules. In particular, we have used these rules
to construct dimensionless quantities
(\textit{e.g.}, $\mu r$, $\omega/\mu$, $\mu M$, $B^\mu/(\mu\sqrt{\mu_0})$, etc.),
used in the numerical implementation and in the results reported in previous sections.
We now proceed to recover units of different physical quantities in order to compare
magnetostatic boson stars with magnetized neutron star solutions.
Restoring $c$ and $G$, the dimensionless quantities related to the total
mass of the star and the norm of the magnetic field ($B^2=B^\mu B_\mu$) are
$Gc^{-2}\mu M$ and $Gc^{-4} B^2/(\mu^2\mu_0)$. Furthermore, the product
\begin{equation}\label{eq:I}
  \mathcal{I}:=\frac{1}{c^4}\sqrt{\frac{G^3}{\mu_0}}\ MB,
\end{equation}
is dimensionless and, more importantly, do not rescale with $\mu$.
Evaluating the magnetic field at the center of the star
we define $\mathcal{I}_c=\mathcal{I}|_{r=0}$ and plot it along $m=1$ sequences
in Fig.~\ref{fig:magB}.

For a neutron star with mass in the range $1.1M_\odot\lesssim M\lesssim 2.1M_\odot$
and strong magnetic fields at the star's pole within the interval
$10^9 \mathrm{T}\lesssim B_\mathrm{pole}\lesssim 10^{11}\mathrm{T}$
\cite{Kaspi:2017fwg}, the value
of the product of mass and magnetic field is between
$10^{-7}\lesssim\mathcal{I}_\mathrm{pole}\lesssim3\times 10^{-5}$.
Internal magnetic fields in magnetars have been estimated according to simulations
to be as high as\footnote{Restricting to the
Einstein-Maxwell-Euler self-consistent models of neutron stars with equations
of state independent of the magnetic field, the
maximum magnetic field $B_c$, which is attained at the center of the star,
is approximately only one order of magnitude bigger than $B_\mathrm{pole}$
(see \textit{e.g.}, \cite{Bocquet:1995je}).}
$10^{14} \mathrm{T}$.
An extended range for $\mathcal{I}$ at the center of
strongly magnetized neutron stars would be
$10^{-7}\lesssim \mathcal{I}_c\lesssim3\times 10^{-2}$.

As can be appreciated from Fig.~\ref{fig:magB}, some of the individual
configurations intersect with the interval
$10^{-7}\lesssim \mathcal{I}_c\lesssim3\times 10^{-2}$, which means
that it is possible to find a value
of $\mu$ such that $B_c$ and $M$ are within the range of neutron stars.
In particular some of the low mass solutions obtained
with $q=1$ are in this region,
while typical, compact solutions with  $q=1,5,25$ and $50$
might have stronger magnetic fields (larger masses)
than magnetars if we assume similar values of the mass (magnetic fields)
of the boson stars to those of the neutron star models.
In order to perform a numerical application we restrict to the
configurations with $q=1$ and $M\mu=0.7$ and choose $M=1.5M_\odot$. This fixes
$\mu(c\hbar)=1.8\times 10^{-10}\mathrm{eV}$ which in turn sets the
magnitude of all other physical variables, for instance the magnitude of the magnetic field at the center of coordinates takes then the value of 
$B_c\approx 1\times 10^{14}\mathrm{T}$, which is 
within the expected values 
of magnetic fields inside magnetars. Furthermore, the size of this bosonic configuration, 
which can be estimated from the size of the torus, is of order $r_{\max(\phi)}\approx 9\ \mathrm{km}$,
obtaining a compactness comparable to those of neutron stars.


\section{Conclusions}
\label{Sec:Conclusions}

In the present work we have constructed magnetized solutions
of boson stars, which are static, axisymmetric, everywhere
regular and asymptotically flat solutions of the
Einstein-Maxwell-Klein-Gordon system characterized by
the mass parameter of the scalar field $\mu$, the azimuthal
harmonic index (winding number) $m$ and the coupling constant $q$.
The configurations consist of two
contrarotating oppositely charged tori, and we have seen that they give rise to an electrically
neutral current that generates a poloidal magnetic field, according
to the observer at rest.

Comparing with the $q=0$ case, which reduces to the toroidal static boson
stars found in \cite{Sanchis-Gual:2021edp}, we obtained that the electromagnetic
field affects the structure of the star and can noticeably change
their  mass and size. Similarly to other boson star models,
in the magnetostatic solutions obtained in this work,
a maximum mass configuration was found for each $q$,
and in all explored cases the sequence of solutions contain
a region of negative binding energy. Regarding the electromagnetic
part, the dipole magnetic moment $\mathcal{M}$ has been obtained and
an important difference is noted with respect to the rotating charged boson star,
namely that the maximum $\mathcal{M}$ configuration does not corresponds with the maximum mass configuration for every $q$, and is shifted towards the zero mass solution ($\omega=\mu$).

We showed that regions of zero magnetic field
form at the inner part of the torus which grow in size
as one considers large enough values of the electromagnetic coupling constant
$q$. On the other hand, it has also been found that
the electromagnetic contribution to the sources  increases in relation
to the scalar field corresponding sources.
For these reasons and since
we have not found any bound for $q$, it would be interesting
to study in a future work, the numerically challenging solutions with $q\gg1$
and also analyze the asymptotic limit $q\rightarrow\infty$.

The magnetic field has been compared to that of
strongly magnetized neutron stars, obtaining that
for similar values of the total mass of the star,
the inner magnetic field is comparable to that of
magnetars for $q\lesssim1$ compact configurations
and greater, for larger values of $q$, making our objects, besides being interesting by their own value, faithful mimickers of neutron stars.
Since toroidal static boson stars with $q=0$ are known to be unstable
we expect that the obtained solutions (at least for small values of $q$) remain unstable, 
however there is a stabilization (and formation) mechanism for neutral multifield boson stars \cite{Sanchis-Gual:2021edp}
which might be applied to the charged scalar field case.
Starting from the conditions in which these magnetized boson stars are stable,
a mechanism for their formation could be devised.
Boson stars are useful entities in strong gravity research,
in particular in dynamical studies and as toy models
of more complex scenarios. In this sense,
the compactness and magnetic field magnitudes of
magnetostatic boson stars motivates the study of
the collapse and the consequent emission in both the
gravitational and electromagnetic channels.
Such collapse dynamics and multimessenger studies will be presented in future works.


\acknowledgments
We thank Juan Carlos Degollado and Olivier Sarbach
for their useful comments in the elaboration of this manuscript.
This work was partially supported by 
DGAPA-UNAM through grants IN110218 and IN105920, by the CONACyT Network Project No. 376127 ``Sombras, lentes y ondas gravitatorias generadas por objetos compactos astrof\'\i sicos".
VJ acknowledge financial support from CONACyT 
graduate grant program.

\appendix

\section{3+1 decomposition of \texorpdfstring{$T_{\mu\nu}$}{T} and the elliptic system of PDEs}
\label{Sec:3+1}
In terms of energy-momentum tensor decomposed into the 3+1 quantities,
\begin{equation}
E=T_{\mu\nu}n^\mu n^\nu;\quad P_\alpha=-n^\nu T_{\mu\nu} \gamma^\nu_\alpha;\quad S_{\alpha \beta}= T_{\mu\nu} \gamma^\mu_\alpha \gamma^\nu_\beta.
\end{equation}
where $n^\alpha=(1/N,0,0,0)$, $\gamma^\alpha_\beta=\delta^\alpha_\beta+n^\alpha n_\beta$ and $N:=e^{F_0}$,
the Einstein equations can be written \cite{Bonazzola:1993zz,Gourgoulhon:1999vx,Grandclement:2014msa}
as the following system of
elliptic equations for the metric coefficients at Eq.(\ref{eq:metric}),

\begin{subequations}\label{eq:elliptic}
  \begin{eqnarray}
  \Delta_3  F_0 &=&4\pi e^{2F_1} \left(E+S\right)  - \partial F_0 \partial\left( F_0 + F_2 \right), \label{eq:nu} \\
\Delta_2 \left[\left(Ne^{F_2} -1\right)r \sin\theta\right] & = &  8 \pi N e^{2F_1+F_2} r \sin\theta \left(S^r_{\ \, r} + 
    S^\theta_{\ \, \theta} \right) ,
    \label{eq:B} \\
\Delta_2\left(F_1  +  F_0\right) & = & 8 \pi e^{2F_1} S^\varphi_{\ \, \varphi} - \partial  F_0 \partial  F_0 ,  \label{eq:A}
\end{eqnarray}
\end{subequations}
where,
\begin{eqnarray}
\Delta_3&:=& \frac{\partial^2}{\partial r^2} + \frac{2}{r}\frac{\partial}{\partial r} + \frac{1}{r^2} \frac{\partial^2}{\partial\theta^2} + \frac{1}{r^2 \tan \theta}\frac{\partial}{\partial \theta}, \\
\Delta_2&:=& \frac{\partial^2}{\partial r^2} + \frac{1}{r}\frac{\partial}{\partial r} + \frac{1}{r^2} \frac{\partial^2}{\partial\theta^2}, \\
\partial f_1 \partial f_2 &:=& \frac{\partial f_1}{\partial r} \frac{\partial f_2}{\partial r} + \frac{1}{r^2}\frac{\partial f_1}{\partial\theta}\frac{\partial f_2}{\partial\theta}.
\end{eqnarray}

The source terms using the ansatz in Eqs.~(\ref{eq:metric}), (\ref{eq:ansatzphi})
and (\ref{eq:ansatzA})
lead to the following expressions:

\begin{eqnarray}
E + S &=& \frac{4}{N^2} \omega^2 \phi^2 - 2\mu^2\phi^2 \label{e:EpS} +\frac{e^{-2(F_1+F_2)}}{\mu_0r^2\sin^2\theta}\partial C\partial C,\\
S^r_{\ \, r} + S^\theta_{\ \, \theta} & = &  2\left[ 
    \frac{\omega^2}{N^2} - \frac{e^{-2F_2}(m-qC)^2}{ r^2 \sin^2\theta} \right]
    \phi^2  - 2\mu^2\phi^2, \label{e:Srr_Sthth}\\
S^\varphi_{\ \, \varphi} & = & 
    \left[ \frac{\omega^2}{N^2} + \frac{e^{-2F_2}(m-qC)^2}{r^2 \sin^2\theta} \right]
    \phi^2 - e^{-2F_1} \partial \phi \partial \phi - \mu^2\phi^2+\frac{e^{-2(F_1+F_2)}}{2\mu_0r^2\sin^2\theta}\partial C\partial C. 
\end{eqnarray}

The two Klein Gordon Eqs., ~(\ref{eq:kg}), and the Maxwell Eq., ~(\ref{eq:maxwell}),
reduce to,

\begin{eqnarray}\label{eq:elliptic2}
  \Delta_3\phi&=&e^{2F_1}\left(\mu^2-\frac{\omega^2}{N^2}\right)\phi-\partial\phi\partial( F_0+F_2)+e^{2F_1-2F_2}\frac{(m-qC)^2\phi}{r^2\sin^2\theta},\\
  \label{eq:elliptic3}
(2\Delta_2-\Delta_3) C&=&-\partial C\partial( F_0-F_2)-2\mu_0qe^{2F_1}(m-qC)\phi^2.
\end{eqnarray}

We have obtained the following number density currents,
\begin{equation}\label{eq:j1}
j_1^\mu=\left(\omega\frac{\phi^2}{N^2},0,0,\frac{e^{-2F_2}(m-qC)\phi^2}{r^2\sin^2\theta}\right),
\end{equation}
and
\begin{equation}\label{eq:j2}
j_2^\mu=\left(\omega\frac{\phi^2}{N^2},0,0,-\frac{e^{-2F_2}(m-qC)\phi^2}{r^2\sin^2\theta}\right),
\end{equation}
which have been inserted in Eq.~(\ref{eq:maxwell}) to obtain Eq.~(\ref{eq:elliptic3}),
and corresponds to the nontrivial remaining Maxwell
equation, $\square A^\varphi-R^\varphi_{\ \varphi}=-\mu_0J^\varphi$.

The Eqs.~(\ref{eq:elliptic}), (\ref{eq:elliptic2}) and (\ref{eq:elliptic3}) make up the
elliptic system of PDEs for the model (\ref{eq:action})
using the ansatz presented at section \ref{Sec:model}.



\bibliography{ref}


\end{document}